\documentclass[12pt,english,american]{article}
\usepackage[T1]{fontenc}
\usepackage[latin9]{inputenc}
\usepackage{geometry}
\usepackage{soul}
\usepackage{cancel}
\usepackage{color}
\usepackage{babel}
\usepackage{amsthm}
\usepackage{amstext}
\usepackage{amssymb}
\usepackage{enumitem}
\usepackage{amsmath}
\usepackage{simplewick}
\usepackage[unicode=true,pdfusetitle,
 bookmarks=true,bookmarksnumbered=false,bookmarksopen=false,
 breaklinks=true,pdfborder={0 0 0},backref=false,colorlinks=true]
 {hyperref}
\hypersetup{linkcolor=blue,citecolor=blue, urlcolor=blue}
\newcommand{\rc}{\color{black}}
\newcommand{\mg}{\color{black}}
\newcommand{\bc}{\color{black}}
\newcommand{\rcc}{\color{black}}
\newcommand{\crr}{\color{black}}
\newcommand{\bcc}{\color{black}}

\newcommand{\m}{n}

\newcommand{\mcF}{\mathcal{F}}

\usepackage{enumitem}		
\theoremstyle{plain}

\theoremstyle{definition}

\theoremstyle{plain}

\theoremstyle{remark}

\theoremstyle{plain}

\theoremstyle{plain}

\newcommand{\mrm}{\mathrm}

\usepackage{txfonts,refstyle,xcolor}

\newref{con}{name = Conjecture\ }
\newref{prop}{name = Proposition\ }
\newref{def}{name = Definition\ }
\newref{sec}{name = Section\ }
\newref{sub}{name = Section\ }
\newref{thm}{name = Theorem\ }
\newref{lem}{name = Lemma\ }
\newref{cor}{name = Corollary\ }
\newref{fig}{name = Figure\ }

\usepackage{bbm}

\usepackage[all]{xy}

\newcommand{\xyR}[1]{%
     \makeatletter
     \xydef@\xymatrixrowsep@{#1}
     \makeatother
}

\newcommand{\xyC}[1]{%
     \makeatletter
     \xydef@\xymatrixcolsep@{#1}
     \makeatother
}

\newcommand{\ncol}[1]{\color{normalcolor}}

\makeatother

\newcommand{\uu}{g}
\newcommand{\hilb}{\mcF}  
\newcommand{\mcE}{E}

\addto\captionsamerican{\renewcommand{\corollaryname}{Corollary}}
\addto\captionsamerican{\renewcommand{\definitionname}{Definition}}
\addto\captionsamerican{\renewcommand{\lemmaname}{Lemma}}
\addto\captionsamerican{\renewcommand{\propositionname}{Proposition}}
\addto\captionsamerican{\renewcommand{\remarkname}{Remark}}
\addto\captionsamerican{\renewcommand{\theoremname}{Theorem}}
\addto\captionsenglish{\renewcommand{\corollaryname}{Corollary}}
\addto\captionsenglish{\renewcommand{\definitionname}{Definition}}
\addto\captionsenglish{\renewcommand{\lemmaname}{Lemma}}
\addto\captionsenglish{\renewcommand{\propositionname}{Proposition}}
\addto\captionsenglish{\renewcommand{\remarkname}{Remark}}
\addto\captionsenglish{\renewcommand{\theoremname}{Theorem}}
\providecommand{\corollaryname}{Corollary}
\providecommand{\definitionname}{Definition}
\providecommand{\lemmaname}{Lemma}
\providecommand{\propositionname}{Proposition}
\providecommand{\remarkname}{Remark}
\providecommand{\theoremname}{Theorem}

\renewcommand{\1}{\!\!\!}

\newcommand{\one}{\mathbf{1}}

\newcommand{\mcB}{\mathcal B}

\newcommand{\hi}{\hat i}
\newcommand{\hj}{\hat j}
\newcommand{\kas}{\kappa_{*}}

\newcommand{\pho}{\mathrm{f}}

\newcommand{\vv}{v}

\newcommand{\g}{\la }

\newcommand{\wt}{\widetilde}

\newcommand{\ti}{\tilde}

\newcommand{\Om}{\Omega}
\newcommand{\ga}{\gamma}

\newcommand{\ka}{\kappa}

\newcommand{\be}{\beta}

\newcommand{\pa}{\partial}

\newcommand{\vp}{\varphi}

\newcommand{\eps}{\varepsilon}
\newcommand{\de}{\delta}

\newcommand{\De}{\Delta}

\newcommand{\out}{+}

\newcommand{\nin}{\noindent}
\newcommand{\si}{\sigma}
\newcommand{\ph}{\phantom}
\newcommand{\h}{\fr{1}{2}}
\newcommand{\nat}{\mathbb{N}}
\newcommand{\hil}{\mathcal{H}}
\newcommand{\om}{\omega}

\newcommand{\supp}{\mathrm{supp}}
\newcommand{\fr}[2]{\frac{#1}{#2}}
\newcommand{\al}{\alpha}
\newcommand{\real}{\mathbb{R}}

\newcommand{\la}{\lambda}
\newcommand{\non}{\nonumber}
\newcommand{\Ga}{\Gamma}

\newcommand{\lan}{\langle}
\newcommand{\ran}{\rangle}
\newcommand{\dGa}{\mrm{d}\Ga}

\def\proof{\noindent{\bf Proof. }}
\def\qed{$\Box$\medskip}

\newtheorem{theoreme}{Theorem } [section]
\newtheorem{proposition}[theoreme]{Proposition}
\newtheorem{lemma}[theoreme]{Lemma}
\newtheorem{definition}[theoreme]{Definition}
\newtheorem{corollary}[theoreme]{Corollary}
\newtheorem{remark}[theoreme]{Remark}
\newtheorem{example}[theoreme]{Example}
\newtheorem{criterion}[theoreme]{Criterion}

\newcommand{\beq}{\begin{equation}}
\newcommand{\eeq}{\end{equation}}
\newcommand{\beqa}{\begin{eqnarray}}
\newcommand{\eeqa}{\end{eqnarray}}
\newcommand{\beqaa}{\begin{align}}
\newcommand{\eeqaa}{\end{align}}
\newcommand{\ben}{\begin{arabicenumerate}}
\newcommand{\een}{\end{arabicenumerate}}
\newcommand{\bex}{\begin{example}}
\newcommand{\eex}{\end{example}}
\newcommand{\ber}{\begin{remark}}
\newcommand{\eer}{\end{remark}}
\newcommand{\bec}{\begin{corollary}}
\newcommand{\eec}{\end{corollary}}
\newcommand{\bep}{\begin{proposition}}
\newcommand{\eep}{\end{proposition}}
\newcommand{\becr}{\begin{criterion}}
\newcommand{\eecr}{\end{criterion}}

\newcommand{\thet}{\gamma}

\def\bel{\begin{lemma}}
\def\eel{\end{lemma}}
\def\bet{\begin{theoreme}}
\def\eet{\end{theoreme}}
\def\bed{\begin{definition}}
\def\eed{\end{definition}}

\begin{document}

\title{Infraparticle states in the massless Nelson model -- revisited} 

\author{Vincent Beaud$^1$,   Wojciech Dybalski$^2$ and Gian Michele Graf$^3$  \\\\
 $^1$Zentrum Mathematik \\ Technische Universit\"at M\"unchen\\
Boltzmannstra\ss e 3, 85748 Garching, Germany\\
\small{E-mail: {\tt vincent.beaud@hotmail.com}}\\\\   
$^2$Faculty of Mathematics and Computer Science \\  Adam Mickiewicz University in Pozna\'n\\
ul. Uniwersytetu Pozna\'nskiego 4, 61-614 Pozna\'n, Poland\\
\small{E-mail: {\tt wojciech.dybalski@amu.edu.pl}} \\\\
$^3$Institut f\"ur Theoretische Physik, ETH Z\"urich \\
Wolfgang-Pauli-Str. 27,  8093 Z\"urich, Switzerland\\
\small{E-mail: {\tt gmgraf@phys.ethz.ch}}
}

\date{}

\maketitle

\begin{abstract}

We provide a new construction of infraparticle states in the massless Nelson model.
The approximating sequence of our infraparticle state does not involve any infrared cut-offs.
{\bcc Its} derivative w.r.t. the time parameter $t$ is given by a simple explicit formula. The convergence
of this sequence as $t\to \infty$ to a non-zero limit is then obtained by the Cook method combined with 
stationary phase estimates. To apply the latter technique we exploit recent results on regularity
of ground states in the massless Nelson model, which hold in the low coupling regime.

\end{abstract}

\section{Introduction}

The massless Nelson model is a time-honoured theoretical laboratory for the infrared aspects of QED. One of its 
variants, which we consider in this work, describes  one non-relativistic massive particle (`the electron'), 
interacting with massless scalar bosons (`the  photons'). The coupling between the electrons and photons 
is chosen in such a way that the model exhibits the \emph{infraparticle problem}, i.e., it does not contain
physical states describing the electron in empty space. In other words, the electron is always accompanied 
by (soft) photons and  it is a challenge to mathematically describe the resulting composite object, usually called
an infraparticle.   {\bcc Two} milestones in rigorous understanding of this problem are works of J. Fr\"ohlich \cite{Fr73,Fr74.1} and A. Pizzo \cite{Pi03,Pi05}.
The latter two papers actually give a complete discussion of the infraparticle in the Nelson model and 
of its collisions with (hard) photons. Also collisions of an infraparticle with a Wigner-type particle (`an atom') in 
a Nelson model with two massive particles are under control \cite{DP19}. However, scattering of several infraparticles
{\bcc appears} steeply difficult in the conventional approach from \cite{Pi05}, as discussed in detail in \cite[Introduction]{DP19}. 
One reason is that  the approximating sequence of the infraparticle state from \cite{Pi05} and the proof of its convergence are technically 
quite intricate,  which {\bc may be due} to limited spectral information on the model available back then. 
Given intervening advances in the spectral theory  \cite{AH12, DP12, DP16}, we revisit the subject and propose  
a simpler approximating sequence of the  infraparticle  in the Nelson model.   {\bcc Its} convergence to a non-trivial 
limit is  relatively straightforward,  given the currently available spectral ingredients.

To explain our construction, let us recall  the definition of the Nelson model.  The Hilbert space of the model is $\hil=L^2(\real^3_x; \mcF)$,
where $\mcF$ is the symmetric Fock space over $L^2(\real^3_k)$. Thus we
will treat $\psi\in \hil$ as  $\mcF$-valued square-integrable functions $\{\psi(x)\}_{x\in \real^3}$, whose scalar product has  the form
\beqa
\lan \psi_1,\psi_2\ran_{\hil}=\int d^3x\, \lan \psi_{1}(x), \psi_{2}(x)\ran_{\mcF}. \label{integral-whole-space}
\eeqa
 The creation and annihilation operators on $\mcF$  are denoted by $a^{(*)}(f)$, $f\in L^2(\real_k^3)$, 
and their sharp variants by $k\mapsto a^{(*)}(k)$.   
The Hamiltonian of the Nelson model has the form
\beqa
H=\fr{(-i\nabla_{x})^2}{2}+H_{\pho}+a(\vv_x)+a^*(\vv_x). \label{H-definition}
\eeqa
Here $x$ and $-i\nabla_x$ are the position and momentum operators on $L^2(\real^3_x)$, 
$(H_{\mrm{f}}, P_{\pho}):=(\dGa(|k|), \dGa(k))$ are the energy-momentum operators  of non-interacting photons and  $\vv_x(k)=\vv(k) e^{-ik\cdot x}$, where  $\vv(k):=\la\fr{\chi_{\ka}(k)}{\sqrt{2|k|}}$ and 
$|\la|\in (0,\la_0]$ is the coupling constant, whose maximal value $\la_0$ will be sufficiently small but non-zero. Here $\chi_{\ka}\in C^{\infty}_0(\real^3)$ is a smooth approximate characteristic function of the ball of radius $\ka=1$\footnote{Although $\ka=1$, it is convenient to keep it in the notation.}. We choose this function rotation invariant, supported in the ball of radius $\ka$ and equal to one on a ball of a slightly smaller radius $(1-\eps_0)\ka$.  By the Kato-Rellich theorem, $H$ is a self-adjoint operator on $D(\h(-i\nabla_{x})^2+H_{\pho} )$. Recalling that the model is translation invariant, 
we denote by $\{H_p\}_{p\in \real^3}$ the usual fiber Hamiltonians acting on the fiber Fock space $\mcF_{\mrm{fi}}$, satisfying
\beqa
H=\Pi^*\big(\int^{\oplus}d^3p\, H_p\, \big)\Pi, \quad \Pi=Fe^{iP_{\mrm{f}} \cdot x},
\eeqa
where $F$ is the Fourier transform in the $x$ variable. In our construction of infraparticle scattering states we will identify the fiber Fock space
$\mcF_{\mrm{fi}}$ with the physical Fock space $\mcF$ which is the reason for the appearance of the unitary $\Pi$ explicitly in formula (\ref{eq:infraparticle}) below.
After this identification, the fiber Hamiltonians are the following self-adjoint operators on $D(P_{\pho}^2+H_{\pho})\subset\mcF$ 
\beqa
H_p:=\h(p-P_{\pho})^2+H_{\pho}+a^*(\vv)+a(\vv), \quad p\in \real^3. \label{fiber-hamiltonian}
\eeqa 
We denote the infimum of the spectrum of $H_p$ by  $E_p$. One manifestation of the infraparticle problem is that $E_p$ is not an eigenvalue. This has been established in considerable generality \cite{Da18, Pi03, Fr74.1}, 
{\bcc  and holds, in particular, for} $ p\in S:=\{\, p'\in\real^3 \,|\, |p'| {\bc <} 1/3\}$ and $\la_0$ sufficiently small. In this range of parameters we know from \cite{AH12} that $p\mapsto E_p$ is real analytic.
It is also well known that  the \emph{modified Hamiltonian} $H_p^{\mrm{w}}$, obtained from  $H_p$ by the
Bogolubov transformation
\beqa
a^{(*)}(k)\mapsto a^{(*)}(k)-f_p(k), \quad f_p(k):=\la\fr{\chi_{\ka}(k)}{\sqrt{2|k|}}\fr{1}{|k|(1-e_k\cdot \nabla E_{p} )}, \quad e_k:=k/|k|, \label{Bogolubov}
\eeqa
is self-adjoint on $D(P_{\pho}^2+H_{\pho})$ and
$E_p$ is its  eigenvalue at the bottom of the spectrum corresponding to an eigenvector $\phi_p$. (Its phase is chosen in the following in accordance with  \cite[Definition 5.2]{DP12}). 

After these preparations we are ready to define the approximating sequences of the infraparticle states.
For any $h\in C^{\infty}_0(\real^3)$ supported in $S$ and any time parameter ${\bcc t\in \real}$  
we set
\begin{align}
\psi_t(x)&:=e^{iHt}e^{-i P_{\pho}\cdot x} {\fr{1}{(2\pi)^{3/2}}}\int d^3p \, e^{i(p\cdot x-E_pt) }  e^{i \thet(p,x,t)}  h(p) 
 W\big( f_p (e^{-i|k|  t+ik\cdot x}-1) \big)  \phi_p,
\label{eq:infraparticle}\\
\thet(p,x,t)&:= \int d^3k f_p(k)^2  \sin(|k|t- k\cdot x ).
\end{align}
Clearly, the Weyl operator $W(g):=e^{a^*(g)-a(g)}$ is well defined for $g(k):=f_p{\rc (k)} (e^{-i|k|  t+ik\cdot x}-1)$, 
 for any {\bc  $(t,x)\in \real^4$}. 
The integral in (\ref{eq:infraparticle}) is well-defined in $\mcF$, since $S\ni p\mapsto \phi_p$ is H\"older continuous
in norm  by  \cite{Pi03} ({\bcc which can also be seen by} \cite[{\bcc formulas~(1.8), (A.4) and Corrolary 5.6}]{DP12} {\bcc combined with Lemma~\ref{energy-bounds-w} below)}.  This integral is an element of  $L^2(\real^3_x;\mcF)$ by Lemma~\ref{time-derivative-lemma}.  Our main result is the following:
\bet\label{lem:time-derivative-final} {\bc There is such $\la_0>0$ that the following holds:} For any $t\in \real$ the vector $\psi_t$ given by (\ref{eq:infraparticle}) belongs to $L^2(\real^3_x;\mcF)$. 
The derivative $\pa_t\psi_t$  exists
in norm in $L^2(\real^3_x;\mcF)$ and we have
\beqa
\pa_t\psi_t	= e^{iHt} e^{-iP_{\pho}\cdot x } \fr{1}{(2\pi)^{3/2}} \int d^3p \,  e^{i (p \cdot x -E_pt  )} e^{i \thet(p,x,t)} i  \ga_{\mrm{int}}(p,x,t)  h(p) W(f_p (e^{-i|k|  t+ik\cdot x }-1))
	\phi_p, \label{eq:time-derivative-final}
\eeqa
where  $\ga_{\mrm{int}}(p,x,t):= 2 \int d^3k\, f_p(k)^2(|k|-k\cdot \nabla E_p) \cos(|k|t-k\cdot x)$ is rapidly decreasing in the region $|x|/t<1$ \\ (cf. Lemma~\ref{phase-lemma-zero}). Furthermore,
\beqa
\int_0^{\infty} dt\, \|\pa_t\psi_t\|_{\hil}<\infty, \label{pat-bound}
\eeqa
hence $\psi^{\out}:=\lim_{t\to\infty} \psi_t$ exists in the norm of $L^2(\real^3_x;\mcF)$. For $h\neq 0$ and ${\bc |\la|\in  (0, \la_0]}$ sufficiently
small, $\psi^{\out}\neq 0$. {\bcc Analogous statements hold for incoming scattering states.}
\eet

The most remarkable part of the theorem is the explicit formula for $\pa_t\psi_t$ given in (\ref{eq:time-derivative-final}). 
It can be anticipated by formal computations on $\mcF$ noting the key relation
\beqa
T(p,x,t)^{*} \left(-i\nabla_x-P_\pho\right)T(p,x,t)= -i\nabla_x - P_\pho^\mrm{w}\quad \textrm{ for } \quad T(p,x,t) :=  W(f_p (e^{-i|k|  t+ik\cdot x }-1)) e^{i \thet(p,x,t)},
 \label{T-relation}
 \eeqa
{\rc where $P_\pho^\mrm{w}$ is obtained from $P_{\pho}$ via the Bogolubov transformation~(\ref{Bogolubov}). Relation~(\ref{T-relation}}) allows to reconstruct  $H^{\mrm{w}}_p$ in front of $\phi_p$ and make use of $H^{\mrm{w}}_p\phi_p=E_p\phi_p$.
It dictates  our  choice of the phase $\ga$ and it is noteworthy that  the resulting $\ga_{\mrm{int}}$
enjoys a rapid decay in $t$ in the physical region of velocities of the electron. This coincidence suggests that our approximating
vector (\ref{eq:infraparticle}) captures optimally the asymptotic dynamics of the Nelson model in the infrared regime.  The decay of $\ga_{\mrm{int}}$
is   {\bcc the} driving force of our convergence
argument based on the Cook method. It also allows for a simple proof of non-triviality of the limit for small $|\la|$. 

Given formula (\ref{eq:time-derivative-final}) and the above remarks,
it may seem very easy to prove the theorem. But it should be kept in mind,  that estimate~(\ref{pat-bound}) must hold in the norm of $L^2(\real^3_x;\mcF)$, which involves the integral over whole space, cf. formula~(\ref{integral-whole-space}) above. To control this integral we use the stationary phase
method,   which generates derivatives w.r.t. $p$ up to the second order   (cf. Lemma~\ref{stationary-phase} below).
Since differentiability of $p\mapsto \phi_p$ is not settled, we have to approximate $\phi_p$ with $\phi_{p,\si}$, which 
come from the Nelson model with an infrared cut-off $\si>0$ in the interaction.  The function $p\mapsto \phi_{p,\si}$ is differentiable
and its derivatives up to the second order have only a mild infrared divergence of the form
\beqa
\|\pa_{p}^{\al} \phi_{p,\si}\|_{\mcF}\leq c\si^{-\de_{\la_0}}, \quad {\rc |\al|=0,1,2,}
\eeqa
where $\de_{\la_0}>0$ tends to zero with $\la_0\to 0$. This estimate, and similar bounds for the wave functions of $\phi_{p,\si}$, 
rely on technical advances from \cite{DP12, DP16}.  Thus, at our present level of understanding, we can eliminate the infrared
cut-off from the formulation of Theorem~\ref{lem:time-derivative-final}, but not from its proof.  

This paper is organized as follows: In Section~\ref{Preliminaries} we provide some technical information, in particular about the model with infrared cut-off. Section~\ref{main-section} is devoted to the proof of Theorem~\ref{lem:time-derivative-final}. In Conclusions we provide a brief comparison of our infraparticle states with the Faddeev-Kulish approach. {\bcc More technical parts of the discussion are postponed to Appendices.}

\vspace{0.2cm}

\noindent{\bf Acknowledgment:} W.D. was partially supported by the  Emmy Noether grant DY107/2-1 of the DFG and its extension DY107/2-2,  and by the NCN grant `Sonata Bis' 2019/34/E/ST1/00053.

\section{Preliminaries}\label{Preliminaries}
\setcounter{equation}{0}

Recall that $\{H_p\}_{p\in \real^3}$ are the fiber Hamiltonians (\ref{fiber-hamiltonian})  and  let $\{H_{p,\si}\}_{p\in \real^3}$ 
be their counterparts at an infrared cut-off $0<\si\leq \ka$. This means that the form factor $\vv$, appearing in (\ref{H-definition}), is replaced with $\vv^{\si}$ 
 given by
\beqa
v^{\si}(k):=\la \fr{\chi_{\rc [\si,\kappa)}(k) }{\rc \sqrt{2|k|} }. \label{form-factor-definitions}
\eeqa
Here $\chi_{[\si,\ka)}(k)=\one_{\mcB'_{\si} }(k)\chi_{\ka}(k)$, $\mcB'_{\si}$
is the complement of the ball of radius $\si$ and $\one_{\De}$ is the characteristic function of    {\bcc a} set 
$\De$. {\rc We remark that  $\{H_{p,\si}\}_{p\in \real^3}$ act on a {\bcc dense} domain in $\mcF$, that is no infrared cut-off is introduced on the Fock space.}
We will work in the range of parameters for which the technical results of \cite{DP12, DP16, DP19}
hold. That is,
\beqa
\la\in  {\bc \mcB_{\la_0}}, \quad \si\in (0,\ka_{\la_0}], \quad p\in S:=\{\, p'\in\real^3 \,|\, |p'|{\bc <} 1/3\},
\eeqa 
where $\la_0$ is sufficiently small and $0<\ka_{\la_0}\leq \ka$. As the fiber  Hamiltonians $H_{p}, H_{p,\si}$  are bounded from below, we can define
\beqa
E_{p}:=\mrm{inf}\,\boldsymbol{\si}(H_p), \quad E_{p,\si}:= \mrm{inf}\,\boldsymbol{\si}(H_{p,\si}),
\eeqa
where $\boldsymbol{\si}$ denotes the spectrum. {\bc (Occasionally we will write $E_p^{(\la)}$, $E_{p,\si}^{(\la)}$ etc. if the dependence on $\la$ will play a role).} $E_p$ enters our definition of the 
 infraparticle state (\ref{eq:time-derivative-final}) and our analysis relies on the following result:
\bel\emph{\cite{AH12}} \label{energy-lemma}  The function  $S\times \mathcal{B}_{\la_0}\ni (p,\la) \mapsto E_{p}^{(\la)}$ is  real-analytic and non-constant. It satisfies $|\nabla E^{{(\la)}}_{p}|\leq 1/2$ and its Hessian matrix {\bc in the $p$-variable}  is strictly positive in $S$ uniformly in $\la$.
\eel
We recall that the modified Hamiltonians $H^{\mrm{w}}_p$ are obtained from $H_p$ by the Bogolubov transformation (\ref{Bogolubov})
and their ground states are denoted $\phi_p$.
Similarly, the modified Hamiltonians $H^{\mrm{w}}_{p,\si}$ are obtained from $H_{p,\si}$ by the transformation
\beqa
a^{(*)}(k)\mapsto a^{(*)}(k)-f_{p,\si}(k), \quad 
f_{p,\si}(k):=\la\fr{\chi_{\rc [\si,\ka)}(k)}{\sqrt{2|k|}}\fr{1}{|k|(1-e_k\cdot \nabla E_{p,\si} )} 
\label{Bogolubov-cut}
\eeqa
and their ground states are denoted $\phi_{p,\si}$. Both $\phi_p$ and $\phi_{p,\si}$ are in the domain of any power of $H_{\pho}$ (cf. Lemma~\ref{energy-bounds-w})) and in addition $\phi_{p,\si}$ are in the domain of any power of the number operator $N:=\dGa(1)$.  For a  choice of the phases of $\phi_p, \phi_{p,\si}$ as in  \cite[Definition 5.2]{DP12}  the following estimate holds 
\beqa
\|(H_{\pho})^{\ell} (\phi_{p}-\phi_{p,\si})\|_{\mcF}\leq c\si^{1/5}, \quad p\in S, \quad \ell\in \nat_0, \label{spectral-bound-intro}
\eeqa
provided that  $\la_0>0$ is readjusted for each $\ell$. 
It is well known for $\ell=0$ \cite{Pi03},  \cite[Corollary 5.6 (a)]{DP19} and for $\ell\in \nat$ it is shown in Appendix~\ref{approximation-appendix}.
{\bcc We will also need the following lemma:}
\bel \label{modified-corollary} 
Let $\ell_1,\ell_2\in \nat_0$.  Then, for $\si\in (0, \ka_{\la_0}]$, 
\beqa
\| H_{\pho}^{\ell_1}N^{\ell_2}\pa_p^{\al} \phi_{p,\si}\|_{\mcF}\leq \fr{c}{\si^{\de_{\la_0} } } \quad \textrm{ for }\quad |\al|=0,1,2. \label{modified-cor-eq}
\eeqa
 The function $\la_0\mapsto \de_{\la_0}$ is positive and satisfies $\lim_{\la_0\to 0} \de_{\la_0}=0$. {\bcc This function and the constant $c$ are}   independent of $p, \si$ within the assumed restrictions,
 but may depend on $\ell_1, \ell_2$.
\eel
\nin {\bcc In Appendix~\ref{first-appendix} we show how to extract the proof of Lemma~\ref{modified-corollary} from \cite{DP12, DP16}.
We remark that  Lemma~\ref{energy-lemma},  bound~(\ref{spectral-bound-intro}), and  Lemma~\ref{modified-corollary}  
are the technical basis for our discussion in the next section. } 

\vspace{0.1cm}

{\bc \nin\textbf{Notation}. {\bcc As we will discuss only outgoing scattering states, we set $t\geq 1$.}  We denote by $c$ numerical constants which may change from line to line. These constants are  independent of $\si$, $p$,  $\la$, $t$,$x$  within the assumed restrictions, but may depend on $\la_0$, 
$\eps_0$. Similarly,  functions denoted $\la_0\mapsto \de_{\la_0}$ are positive {\bc and satisfy}  $\lim_{\la_0\to 0} \de_{\la_0}=0$. They are independent of $\si$, $p$ within the assumed restrictions but may depend on $\eps_0$. These functions may change from line to line.}

\vspace{0.1cm}

\section{Infraparticle states} \label{main-section}
\setcounter{equation}{0}

The goal of this section is to provide a proof of Theorem~\ref{lem:time-derivative-final}.
Our main tool  will be the stationary phase method. The estimates suitable for our purposes are stated in the following lemma, which is proven  in  Appendix~\ref{A-appendix}. 

\bel\label{stationary-phase}  Let $p\mapsto \uu(p)\in \hilb$ be  weakly infinitely differentiable on some dense domain 
and compactly supported in $S$.   Let $c_0$ be s.t. $|\nabla \mcE_p|<c_0<1$ for $p\in \supp\,{\rc \uu}$.    
Then, for   any $0\leq \eps\leq 1/2$,
\beqa
\bigg( \int_{|x|/t\leq c_0} d^3x\,\| \int d^3p \, e^{i(p\cdot x-{\bc \mcE_p} t)} \uu(p)\|_{\hilb}^2 \bigg)^{1/2}\1 &\leq&\1 {\rcc c
 \sum_{|\al|\leq 2 }\sup_{p,|x|\leq c_0t} \| \pa_{p}^{\al} \uu(p)\|_{\hilb}}, 
 \label{stationary-phase-small-vel}\\
\bigg( \int_{|x|/t\geq c_0} d^3x\,\| \int d^3p \, e^{i(p\cdot x-\mcE_p t)} \uu(p)\|_{\hilb}^2 \bigg)^{1/2} \1&\leq &\1
{\rcc c t^{-1/2+\eps}  \sum_{|\al|\leq 2 } \sup_{p,|x|\geq c_0t} \bigg(\fr{1}{(1+|t|+|x|)^{\eps}}    \|\pa_{p}^{\al} \uu(p)\|_{\hilb}\bigg).}
\quad\quad \label{stationary-phase-large-vel}
\eeqa 
{\rc The function $\uu$ above may depend on $(x,t)$.}
\eel
\nin Lemma~\ref{stationary-phase}  immediately gives the following estimate 
\beqa
{\rcc \bigg( \int d^3x\,\| \int d^3p \, e^{i(p\cdot x- \mcE_p t)} \uu(p)\|_{\hilb}^2 \bigg)^{1/2}  } \leq 
c{\bc   t^{1/2} }\sum_{|\al|\leq 2 }\sup_{p, x} \bigg(\fr{1}{(1+|x|)^{{\bc 1/2}}}    \| \pa_{p}^{\al} \uu(p)\|_{\mcF}\bigg),
\label{finite-time-estimate}
\eeqa
which will be useful for analyzing  vectors (\ref{eq:infraparticle}) at finite $t$. We note that we
cannot apply (\ref{finite-time-estimate}) or Lemma~\ref{stationary-phase} directly to the infraparticle vector 
(\ref{eq:infraparticle}), since differentiability of $p\mapsto \phi_p$ is out of control. In the course 
of our discussion we will approximate $\phi_p$ with $\phi_{p,\si}$ in a suitable manner.

As a first step of our analysis, we compute and estimate derivatives of  $e^{i \thet(p,x,t)}$ w.r.t. $p,x,t$. The following
is a result of a straightforward computation: 
\beqa
\pa_t e^{i \thet(p,x,t)} \1&=&\1 e^{i \thet(p,x,t)} {\bc i}  \int d^3k\, f_p(k)^2|k|  \cos(|k|t- k\cdot x ),\label{first-formula} \\
\pa_{x_i} e^{i \thet(p,x,t)}  \1&=&\1  -e^{i \thet(p,x,t)} {\bc i}  \int d^3k\, f_p(k)^2  k_i \cos(|k|t- k\cdot x ), \\
\pa_{x_j}\pa_{x_i} e^{i \thet(p,x,t)}  \1&=&\1 {\bc-}e^{i \thet(p,x,t)} \int d^3k\, f_p(k)^2  k_j \cos(|k|t- k\cdot x )  \int d^3k f_p(k)^2  k_i \cos(|k|t- k\cdot x )  \\
& &-e^{i \thet(p,x,t)} {\bc i}  \int d^3k f_p(k)^2  k_ik_j \sin(|k|t- k\cdot x ). \label{last-formula}
\eeqa
Now we estimate the above expressions together with their derivatives w.r.t. $p$.
\bel\label{gamma-bounds} The following bounds hold
\beqa
|\pa^{\al}_p \pa^{\ell}_t e^{i \thet(p,x,t)}| \1&\leq&\1 c(1+\log(1+|t|+|x|))^2,\\
|\pa^{\al}_p\pa^{\be}_x e^{i \thet(p,x,t)}| \1&\leq&\1 c(1+\log(1+|t|+|x|))^2,
\eeqa
for $|\al|, |\be| \leq 2$, $\ell\leq 1$.
\eel
\proof We see from (\ref{first-formula})--(\ref{last-formula}) that the derivatives w.r.t. $x,t$ produce expressions which are
uniformly bounded in $x,t$ due to the additional factors $k_i, |k|$, which regularize the singularity of $f_p^2$ {\bc at $|k|=0$}. 
Hence, it suffices to study the expression
\beqa
\pa_{p_j}\pa_{p_i}  e^{i \thet(p,x,t)} \1&=&\1  \pa_{p_j}\big( e^{i \thet(p,x,t)} i \pa_{p_i}\thet(p,x,t) \big)\non\\
\1&=&\1  e^{i \thet(p,x,t)} \big(i \pa_{p_j}\thet(p,x,t)\big)  \big(i \pa_{p_i}\thet(p,x,t)\big) + e^{i \thet(p,x,t)} i \pa_{p_j}\pa_{p_i}\thet(p,x,t).
\eeqa
Making use of (\ref{f_p-zero-estimate}), we obtain
\beqa
|\pa_{p_j}\pa_{p_i}  e^{i \thet(p,x,t)}|\leq c(1+\log(1+|t|+|x|))^2,
\eeqa
{\rc where the dependence of $c$ on parameters is as discussed in Section~\ref{Preliminaries}.}  This concludes the proof. \qed\\
As a next step of our discussion we compute derivatives of the following auxiliary vector
\beqa
\hat{\uu}_{(t,x)}(p)=W\big( f_p m(t,x) \big)  \phi_p, \quad m(t,x):=u(t,x)-1, \quad u(t,x):=e^{-i|k|  t+ik\cdot x} \label{hat-vectors}
\eeqa
w.r.t. $(t,x)$ up to the second order. {\bc We will abbreviate $m:=m(t,x), u:=u(t,x)$}. 
\bel The function $(t,x)\mapsto {\rc \hat{\uu}}_{(t,x)}(p)$ is infinitely often partially differentiable in the norm of $\mcF$ and
the following formulas hold
\beqa
\pa_t\hat{\uu}_{(t,x)}(p)\1&=&\1  W\big( f_p m \big)i\big(\Phi( f_p \pa_{t}m)+\mrm{Im}\lan f_p m, f_p \pa_{t} m\ran \big) \phi_p,  \\
\pa_t^2\hat{\uu}_{(t,x)}(p)\1&=&\1-W\big( f_p m \big) \big(\Phi( f_p \pa_{t}m)+\mrm{Im}\lan f_p m, f_p \pa_{t} m\ran \big)^2 \phi_p
+W\big( f_p m \big) i\big(\Phi( f_p \pa_{t}^2m)+  \mrm{Im}\lan f_p m, f_p \pa_{t}^2 m\ran \big) \phi_p,\quad\\
\pa_{x_i}\hat{\uu}_{(t,x)}(p)\1&=&\1 W\big( f_p m \big)i\big(\Phi( f_p \pa_{x_i}m)+\mrm{Im}\lan f_p m, f_p \pa_{x_i} m\ran \big) \phi_p, \\
\pa_{x_j}\pa_{x_i}\hat{\uu}_{(t,x)}(p)\1&=&\1 W\big( f_p m \big) i\big(\Phi( f_p \pa_{x_j}m)+\mrm{Im}\lan f_p m, f_p \pa_{x_j} m\ran \big)  i\big(\Phi( f_p \pa_{x_i}m)+\mrm{Im}\lan f_p m, f_p \pa_{x_i} m\ran \big) \phi_p\non\\
\1& &\1+W\big( f_p m \big) i\big(\Phi( f_p \pa_{x_j}\pa_{x_i}m)+\mrm{Im}\lan f_p \pa_{x_j}m, f_p \pa_{x_i} m\ran 
+ \mrm{Im}\lan f_p m, f_p \pa_{x_j}\pa_{x_i} m\ran \big) \phi_p, \label{two-x-derivatives}
\eeqa
where $\Phi(F):=a^*(-iF)+a(-iF)$, $F\in L^2(\real^3_k)$. 
\eel
\proof   We note that, by Lemma~\ref{energy-bounds-w}, $\phi_p$ belongs to  $D(H_{\pho}^{\ell})$ for any $\ell\in \nat$.
  We {\bc observe} that {\bcc for any fixed $(t,x)$ the function} $f_p m(t,x)\in L^2_{\om}(\real^3_k)$  and it is infinitely differentiable  in $(t,x)$  in the norm of $L^2_{\om}(\real^3_k)$ {\bc (see Appendix~\ref{Weyl-differentiability})}.  For the first derivative w.r.t. $x_i$ this follows from 
 \beqa
m(t, x+(\De x)_i{\bcc e_i})=  m(t,x)+(\De x)_i (\pa_{x_i} m)(t,x)+  (\De x)_i^2 \int_0^{1}ds\,(1-s)\,  (\pa_{x_i}^2 m)(t, x+s(\De x)_i{\bcc e_i}), \label{Taylor}
\eeqa
and the fact that $|k|^{-1}\pa_{x_i}^{\ell}m(t,x)$ is bounded in $k$ for any $\ell\in \nat_{\bcc 0}$. For higher derivatives we simply
replace $m$ with ${\rc \pa_{x_i}^{\ell}m}$ in (\ref{Taylor}). The arguments regarding the derivatives w.r.t. $t$ are analogous. 
Thus we can compute the derivatives using Lemma~\ref{W-der-lemma},
which gives the formulas from the statement of the lemma. \qed\\
Now we analyze the regularized variants of the vectors from (\ref{hat-vectors})
\beqa
\hat{\uu}^{\si}_{(t,x)}(p):=W\big( f_p m(t,x) \big)  \phi_{p,\si}.
\eeqa
We note  the following {\bc fact}:
\bel\label{u-bounds} There hold the  bounds
\beqa
\|\pa^{\al}_p \pa_t^{\ell}  \hat{\uu}^{\si}_{(t,x)}(p)\|_{\mcF} \1&\leq&\1c \fr{(1+\log(1+|x|+|t|))^{  {\bcc  3}  }}{\si^{\de_{\la_0}}}, \label{many-derivatives-zero} \\
\|\pa^{\al}_p\pa^{\be}_x\hat{\uu}^{\si}_{(t,x)}(p)\|_{\mcF} \1&\leq&\1c \fr{(1+\log(1+|x|+|t|))^{{\bcc \,\, 3}} }{\si^{\de_{\la_0}}}, \label{many-derivatives}
\eeqa
for $\ell, |\al|, |\be|\leq 2$ {\crr and $\si\in (0,\ka_{\la_0}]$}. The $x$ and $t$ derivatives exist in the norm of $\mcF$. The derivatives w.r.t. $p$ exist in {\bc the} weak sense on
{\bc the domain of  finite particle vectors with compactly supported wave functions (cf. \cite[p. 208]{RS2})}. 
{\bc The bound (\ref{many-derivatives}) still holds if $\pa^{\be}_x$ is replaced with $H_{\pho}, P_{\pho,i}, P_{\pho,i}^2$  or $\pa_{x_i}P_{\pho,i}$. } 
\eel
\proof We consider only (\ref{many-derivatives}) for $|\al|=2, |\be|=2$ as the remaining cases are analogous and simpler.
 To handle the resulting expressions, 
it is convenient to define, for $s\mapsto F_s$ as in Lemma~\ref{W-der-lemma},
\beqa
\wt{\Phi}_s(F):=\Phi(\pa_s F_s)+\mrm{Im}\lan F, \pa_s F_s\ran.
\eeqa
Using this notation and recalling  (\ref{two-x-derivatives}), we can write  
\beqa
\pa_{x_j}\pa_{x_i}\hat{\uu}^{\si}_{(t,x)}(p)\1&=&\1 W\big( f_p m \big) \bigg\{ i\wt\Phi_{x_j}( f_p m) i\wt\Phi_{x_i}( f_p m)
+ i  \pa_{x_j} \wt\Phi_{x_i}( f_p  m)\bigg\} \phi_{p,\si}
= W\big( f_p m \big)P_{x_i,x_j}(f_pm) \phi_{p,\si},
\eeqa
where in the last step we denoted the expression in curly bracket by $P_{x_i,x_j}(f_pm)$ to further abbreviate the 
{\bcc notation}. 
Now we compute the first derivative w.r.t. momentum. We recall that these derivatives must only exist weakly 
{\bc on the domain of finite particle vectors},
i.e., after taking
a scalar product with {\bc such vectors}. 
{\bc This will control the unbounded operators acting on $\phi_{p,\si}$ below and, in particular,} allow us to differentiate $p\mapsto \phi_{p,\si}$ {\bc in (\ref{first-derivative-three}) below}.  In this sense, we compute: 
\beqa
\pa_{p_{\hi}} \pa_{x_j}\pa_{x_i}\hat{\uu}^{\si}_{(t,x)}(p)
\1&=&\1W\big( f_p m \big)i \wt{\Phi}_{ p_{\hi} }(f_p m) P_{x_i,x_j}(f_pm)  \phi_{p,\si} \label{first-derivative-one} \\
\1& &\1+ W\big( f_p m \big) \pa_{p_{\hi}}  \big(P_{x_i,x_j}(f_pm)\big)  \phi_{p,\si} \label{first-derivative-two} \\
\1& &\1+ W\big( f_p m \big)   P_{x_i,x_j}(f_pm) \pa_{p_{\hi}}\phi_{p,\si}. \label{first-derivative-three}
 \eeqa
Now we compute the respective contributions to $\pa_{p_{\hj}} \pa_{p_{\hi}} \pa_{x_j}\pa_{x_i}{\rc \hat{\uu}}^{\si}_{(t,x)}(p)$:
(\ref{first-derivative-one}) gives
\beqa
\pa_{p_{\hj}} \big(W\big( f_p m \big)i \wt{\Phi}_{p_{\hi}}(f_p m) P_{x_i,x_j}(f_pm)  \phi_{p,\si}\big)\1&=&\1
W\big( f_p m \big)i \wt{\Phi}_{p_{\hj}}(f_p m) i \wt{\Phi}_{p_{\hi}}(f_p m)  P_{x_i,x_j}(f_pm)  \phi_{p,\si} \label{highest-order}\\
\1& &\1+W\big( f_p m \big)i  \pa_{p_{\hj}} (\wt{\Phi}(f_p m) P_{x_i,x_j}(f_pm))  \phi_{p,\si} \\
\1& &\1+W\big( f_p m \big)i \wt{\Phi}(f_p m) P_{x_i,x_j}(f_pm)  \pa_{p_{\hj}}\phi_{p,\si}.
\eeqa
From (\ref{first-derivative-two}) we obtain
\beqa
\pa_{p_{\hj}} \big(W\big( f_p m \big) \pa_{p_{\hi}}  P_{x_i,x_j}(f_pm)  \phi_{p,\si}\big)
\1&=&\1 W\big( f_p m \big) i\wt{\Phi}_{p_{\hi}}(f_p m) \pa_{p_{\hi}}  P_{x_i,x_j}(f_pm)  \phi_{p,\si}\non\\
\1& &\1+ W\big( f_p m \big) \pa_{p_{\hj}}\pa_{p_{\hi}}  \big(P_{x_i,x_j}(f_pm)\big)  \phi_{p,\si} \non\\
\1& &\1+W\big( f_p m \big) \pa_{p_{\hi}}  P_{x_i,x_j}(f_pm) \pa_{p_{\hj}} \phi_{p,\si}.
\eeqa
From (\ref{first-derivative-three}) we get
\beqa
\pa_{p_{\hj}} \big( W\big( f_p m \big)   P_{x_i,x_j}(f_pm) \pa_{p_{\hi}}\phi_{p,\si}\big) \1&=&\1 W\big( f_p m \big) i\wt{\Phi}_{p_{\hi}}(f_p m) \  P_{x_i,x_j}(f_pm) \pa_{p_{\hi}} \phi_{p,\si}\non\\
\1& &\1+ W\big( f_p m \big) \pa_{p_{\hj}}  \big(P_{x_i,x_j}(f_pm)\big)  \pa_{p_{\hi}}\phi_{p,\si} \non\\
\1& &\1+ W\big( f_p m \big)   P_{x_i,x_j}(f_pm) \pa_{p_{\hi}}\pa_{p_{\hj}} \phi_{p,\si}.
\eeqa
To estimate these expressions, we recall from Lemma~\ref{modified-corollary} that $\pa^{\al}_p \phi_{p,\si}$ are in the 
domain of any power of $N$ and $\|N^{\ell} \pa^{\al}_p \phi_{p,\si}\|_{\mcF} \leq c_{\ell} \si^{-\de_{\la_0}}$. Thus making use
of the number bounds (\ref{number-bounds-eq}), we have
\beqa
\|\pa_{p_{\hj}} \pa_{p_{\hi}} \pa_{x_j}\pa_{x_i}\hat{\uu}^{\si}_{(t,x)}(p)\|_{\mcF}\leq  P( \|f_pm\|_2, \|f_p\pa_{x_i}m\|_2,   \|f_p \pa_{x_j}\pa_{x_i}m\|_2)  \si^{-\de_{\la_0}}. \label{polynomial-bound}
\eeqa 
Here $P$ is a certain polynomial in the specified norms, which also includes  $\|\pa_{p}^{\al}f m\|_2$.
We recall, however, that $f_p(k):=\la\fr{\chi_{\ka}(k)}{\sqrt{2|k|}}\fr{1}{|k|(1-e_k\cdot \nabla E_{p} )}$, thus derivatives of $f_p$ w.r.t.
$p$ only  change  the behaviour of this function in the angular variable $e_k$ but not in the $|k|$-variable. As our estimates
are insensitive to the angular behaviour, we omitted  {\bcc these derivatives} in the notation in (\ref{polynomial-bound}). Making use of Lemma~\ref{F-lemma},
we have
\beqa
\|f_pm\|_2\leq c|\la|(1+\log(1+|x|+|t|))^{1/2}, \quad \|f_p\pa_{x_i}m\|_2\leq c|\la|, \quad  \|f_p\pa_{x_i}\pa_{x_j}m\|_2\leq c|\la|. \label{f-p-m}
\eeqa
By inspection, we see that $P$ is at most of  the {\bcc sixth}  order in $\|f_pm\|_2$ (cf. (\ref{highest-order})), which concludes the proof of estimates 
(\ref{many-derivatives-zero}), (\ref{many-derivatives}).

{\bc As for the last statement of the lemma,  the case of $H_{\pho}, P_{\pho,i}, P_{\pho,i}^2$ is covered by the fact that the derivatives
w.r.t. $p$ should exist only weakly on vectors which belong to domains of these operators. After computing these derivatives
one pulls $H_{\pho}, P_{\pho,i}, P_{\pho,i}^2$ to the right through the Weyl operator according to
\beqa
H_{\pho}W(f_{p}m)= W(f_{p}m)(H_{\pho}+a^*({\bc |k|}f_pm)+a({\bc |k|}f_pm) +\|{\bc |k|^{1/2}} f_p m \|_2^2) \label{pull-through}
\eeqa
and applies   Lemmas~\ref{modified-corollary} {\bcc and} \ref{energy-bounds-lemma}. The case of $P_{\pho,i} \pa_{x_i}$ requires more consideration as the derivative
w.r.t. $x_i$ should exist in the norm of $\mcF$. To check that $P_{\pho,i}W(f_{p}m)\phi_{p,\si}$ is partially differentiable w.r.t. $x$ in the
norm of $\mcF$, we write, analogously {\bcc to} (\ref{pull-through}),
\beqa
P_{\pho,i}W(f_{p}m)\phi_{p,\si}= W(f_{p}m)(P_{\pho,i}+a^*(k_if_pm)+a(k_i f_pm ) +\lan f_p, k_i f_p\ran ) \phi_{p,\si}
\label{H-W-eq}
\eeqa
and refer to Lemma~\ref{W-der-lemma}. By a computation we obtain 
\beqa
\pa_{x_i} P_{\pho,i}W(f_{p}m )\phi_{p,\si}=  P_{\pho,i} \pa_{x_i}\big(  W(f_{p}m)\big)\phi_{p,\si}, \label{H-W-eq-one}
\eeqa
where $\pa_{x_i}\big(  W(f_{p}m)\big)$ is the explicit formula from Lemma~\ref{W-der-lemma}, and then  proceed as in the discussion of $H_{\pho}, P_{\pho,i}, P_{\pho,i}^2$ above. }  \qed

Now we are ready to analyze the infraparticle vector  (\ref{eq:infraparticle}). 
\bel\label{time-derivative-lemma} {\bc There is such $\la_0>0$ that} for {\bcc any $|\la|\in (0,\la_0]$ and } $t\in \real$, the integral
\beqa
\Psi_t(x):=\int d^3p \, e^{i(p\cdot x-E_pt) }  e^{i \thet(p,x,t)}  h(p) W\big( f_p m(t,x)  \big)   \phi_p \label{vector-Psi}
\eeqa
has the following properties:
\begin{enumerate}
\item[(a)] $\Psi_t\in L^2(\real^3_x;\mcF)$.
\item[(b)] $\Psi_t$ is differentiable in $t$ {\bc in} the norm of $L^2(\real^3_x;\mcF)$ and
\beqa
\pa_t\Psi_t(x)\1&=&\1\int d^3p \, e^{i(p\cdot x-E_pt) } \big(-iE_p+i\pa_t \thet(p,x,t)+i\mrm{Im}\lan f_pm, f_p\pa_t m\ran \big)  e^{i \thet(p,x,t)}  h(p) W\big( f_p m \big)   \phi_p\non\\
\1& &\1+ \int d^3p \, e^{i(p\cdot x-E_pt) }   e^{i \thet(p,x,t)}  h(p) W\big( f_p m \big)(a^*(f_p\pa_t m)-a(f_p\pa_tm))   \phi_p. \label{time-derivative}
\eeqa
\end{enumerate}
\eel
\proof  As for (a),  to prove that $x\mapsto \Psi_t(x)$ is square integrable, we intend to 
apply Lemma~\ref{stationary-phase}. {\bc However}, we lack information about the differentiability of $p\mapsto \phi_p$.
To circumvent this problem, we introduce an  $x${\bc-}dependent cut-off 
$\si_{x}:=\ka_{\la_0}/(1+|x|)^M$, where $M$ is sufficiently large but fixed. We insert  {\bc into (\ref{vector-Psi})}
\beqa
\phi_p=(\phi_p-\phi_{p,\si_{x}})+\phi_{p,\si_{x}} \label{shift-zero}
\eeqa
 and obtain
\beqa
\Psi_t(x)=\int d^3p \, e^{i(p\cdot x-E_pt) }  e^{i \thet(p,x,t)}  h(p) W\big( {\bc f_p m}   \big)   (\phi_p - \phi_{p,\si_{x}})+\Psi_t^{\si_{x}}(x). \label{shift}
\eeqa
Here  $\Psi_t^{\si_{x}}(x)$ is given by (\ref{vector-Psi}) with $\phi_p$ replaced with 
$\phi_{p,\si_{x}}$. Concerning the first term on the r.h.s. of (\ref{shift}), we have by (\ref{spectral-bound-intro}) 
\beqa
\big\|\int d^3p \, e^{i(p\cdot x-E_pt) }  e^{i \thet(p,x,t)}  h(p) W\big( {\bc f_p m}  \big)   (\phi_p - \phi_{p,\si_{{\bc x}}})\big\|_{\mcF}\leq \fr{c (\ka_{\la_0})^{ 1/5 } }{(1+|x|)^{{\bc M}/5} }.
\eeqa
Thus this term is manifestly in $L^2(\real^3_x;\mcF)$ {\bc for $2M/5>3$}. As for the last term on the r.h.s. of (\ref{shift}),
estimate~(\ref{finite-time-estimate}) gives 
\beqa
\|\Psi_t^{\si_{x}}\|_{\hil}\leq c {\bc t^{1/2} }  \sum_{|\al|\leq 2 } \sup_{p,x} \bigg(\fr{1}{(1+|x|)^{{\bc 1/2}}} \|
 \pa_{p}^{\al} (e^{i \thet(p,x,t)} \hat{{\rc \uu}}^{\si_x}_{(t,x)}(p))\|_{\mcF}\bigg). \label{L-2-bound}
 \eeqa
{\bc The expression on the r.h.s. above is finite for any fixed $t$}  by Lemmas~\ref{gamma-bounds}, \ref{u-bounds}, 
{\bc provided $\de_{\la_0}$ of Lemma~\ref{u-bounds} satisfies $M\de_{\la_0}<1/2$. } This concludes the proof of (a).

Part (b) is a straightforward computation, provided we can show differentiability in the norm of $L^2(\real^3_x; \mcF)$. 
To this end, we {\bc use} the Taylor theorem (cf. formula (\ref{Taylor})) 
\beqa
& &\int d^3p \, e^{i(p\cdot x-E_pt) }  e^{i \thet(p,x,t)}  h(p) \bigg( \fr{W\big( f_p m(t+\De t,x) \big) -  W\big( f_p m(t,x) \big)}{\De t}-
 \pa_{t}W\big( f_p m(t,x) \big)  \bigg) \phi_p\non\\
& &\ph{444444444444}= \De t \int d^3p \, e^{i(p\cdot x-E_pt) }  e^{i \thet(p,x,t)}  h(p)\int_0^1 ds\,(1-s) 
{\bc \big\{\pa_{\tau}^2W\big( f_p m(\tau,x) \big)|_{\tau=t+s\De t }\big\}} \phi_p.
 \label{De-integral}
\eeqa
Since $\phi_p$ is in the domain of any power of $H_{\pho}$ (cf. Lemma~\ref{energy-bounds-w}), we can compute  {\bc $\pa_{\tau}^2W\big( f_p m(\tau,x) \big)$}
using Lemma~\ref{W-der-lemma}. Next, exploiting {\bc the} energy bounds (\ref{energy-bounds-eq}) to control the creation and annihilation operators
acting on $\phi_p$, we apply the shift (\ref{shift-zero}) and  estimate (\ref{spectral-bound-intro}). Then, proceeding analogously as in part (a),
we show that  (\ref{De-integral}) tends to zero with  $\De t\to 0$ in the norm of $L^2(\real^3_x;\mcF)$.  Differentiability {\bc in the  norm of $L^2(\real^3_x;\mcF)$ }of other ingredients of (\ref{vector-Psi}) can be shown by analogous and simpler arguments. Now formula (\ref{time-derivative})  follows
by an application of Lemma~\ref{W-der-lemma}. \qed 
\bel\label{Domain-of-H}  Vectors $\Psi_t\in L^2(\real^3_x;\mcF)$, $t\in \real$, defined in~(\ref{vector-Psi}) have the following properties:
\begin{enumerate}
\item[(a)] $\Psi_t$ is in the domain of  $P_{\mrm{f},i}, P_{\pho,i}^2$, $H_{\pho}$ and the following formula holds
\beqa
& &(H_{\mrm{f}} \Psi_t)(x)=\int d^3p \, e^{i(p\cdot x-E_pt) }  e^{i \thet(p,x,t)}  h(p) W\big( f_p m  \big)\big(H_{\pho}^{\mrm{w}}+a^*(|k|f_p u)+a(|k|f_pu)\non\\
& &\ph{44444444444444444444444444444444444444444}\,+{\bcc \lan f_p,|k| f_p\ran}-2\mrm{Re}\lan f_p, |k| f_p u\ran  \big)   \phi_p.
\eeqa

\item[(b)] $\Psi_t$ is in the domain of $-i\pa_{x_i}$, $(-i\pa_{x_i})^2$,  $-i\pa_{x_i}P_{\mrm{f},i}$ and  the following formula holds
\beqa
(-i\pa_{x_i}- P_{\pho,i})^{2} \Psi_t(x)= \int d^3p \, e^{i(p\cdot x-E_pt) }  e^{i \thet(p,x,t)}  h(p) W\big( f_p m \big) \big(p_i-P_{\pho,i}^{\mrm{w}} \big)^{2} \phi_p. 
\eeqa

\item[(c)]    $\Psi_t$ is in the domain of $(a^*(v)+a(v))$ and the following formula holds
\beqa
(a^*(v)+a(v)) \Psi_t(x)=\int d^3p \, e^{i(p\cdot x-E_pt) }  e^{i \thet(p,x,t)}  h(p) W\big( f_p m \big) ( (a^*(v)+a(v))^{\mrm{w}} + 2\mrm{Re}\lan f_pu, v\ran
 ) \phi_p.
\eeqa

\end{enumerate}
\eel
\proof We start with some computations on  $\mcF$ which are justified by Lemma~\ref{W-der-lemma}. Since $W\big( f_p m  \big)\phi_p$ is in the domain of $H_{\pho}$, we can write for any fixed $t$
\beqa
H_{\pho}\Psi_t(x) \1 &=& \1 \int d^3p \, e^{i(p\cdot x-E_pt) }  e^{i \thet(p,x,t)}  h(p) H_{\pho}W\big( f_p m  \big)   \phi_p \non\\
\1&=&\1 \int d^3p \, e^{i(p\cdot x-E_pt) }  e^{i \thet(p,x,t)}  h(p) W\big( f_p m  \big)\big(H_{\pho}{+}a^*(|k|f_p m){+}a(|k|f_pm)+\||k|^{1/2} f_pm\|_2^2\big)   \phi_p\non\\
\1&=&\1  \int d^3p \, e^{i(p\cdot x-E_pt) }  e^{i \thet(p,x,t)}  h(p) W\big( f_p m  \big)\big(H_{\pho}^{\mrm{w}}+a^*(|k|f_p u)+a(|k|f_pu)+
\lan f_p,|k| f_p\ran-2\mrm{Re}\lan f_p, |k| f_p u\ran  \big)   \phi_p, \quad\quad\,\,\,\,
\eeqa
where we made use of {\bcc $H_{\pho}^{\mrm{w}}=H_{\pho}-a^*(|k|f_p)-a(|k|f_p)+\|\, |k|^{1/2}f_p\|_2^2$ }and
\beqa
-\| |k|^{1/2} f_p   \|_2^2 +\||k|^{1/2} f_pm\|_2^2= \lan f_p,|k| f_p\ran -2\mrm{Re}\lan f_p, |k| f_p u\ran.
\eeqa
Analogously, we obtain for ${\bc \ell\in \{1,2\} }$,
\beqa
(P_{\pho,i})^{\ell} \Psi_t(x) \1&=&\1  \int d^3p \, e^{i(p\cdot x-E_pt) }  e^{i \thet(p,x,t)}  h(p) 
W\big( f_p m  \big)\times \non\\
& &\ph{444444444444}\times\big(P^{\mrm{w}}_{\pho,i}+a^*(k_if_p {\bc u}) +a(k_if_pu)+  \lan f_p,k_i f_p\ran -2\mrm{Re}\lan f_p, k_i f_p u \ran\big)^{\ell}   \phi_p. \label{P-eq}
\eeqa
Furthermore, {\bc by similar considerations as in (\ref{De-integral}), we can exchange $-i\pa_{x_i}$ with the $p$-integral}
and obtain the following
\beqa
-i\pa_{x_i}\Psi_t(x)\1&=&\1\int d^3p \, e^{i(p\cdot x-E_pt) } e^{i \thet(p,x,t)}  h(p)\big(p_i+\pa_{x_i} \thet(p,x,t)+\mrm{Im}\lan f_pm, f_p\pa_{x_i} m\ran \big)  
 W\big( f_p m \big)   \phi_p  \non\\
\1& &\1+ \int d^3p \, e^{i(p\cdot x-E_pt) }   e^{i \thet(p,x,t)}  h(p) W\big( f_p m \big)( a^*(k_if_p  u)+a(k_if_p  u))   \phi_p. \label{pax-eq}
\eeqa
Combining the above computations we also obtain
\beqa
-i\pa_{x_i}P_{\pho,i} \Psi_t(x) \1&=&\1  \int d^3p \, e^{i(p\cdot x-E_pt) } e^{i \thet(p,x,t)}  h(p)  \big(p_i+\pa_{x_i} \thet(p,x,t)+\mrm{Im}\lan f_pm, f_p\pa_{x_i} m\ran \big)   \times \non\\
& &\ph{44444444444}\times W\big( f_p m  \big)\big(P_{\pho,i}{+}a^*(k_if_p m) {+}a(k_if_pm)+\lan f_pm, k_i f_pm \ran   \big)   \phi_p\non\\ 
\1&+&\1  \int d^3p \, e^{i(p\cdot x-E_pt) }  e^{i \thet(p,x,t)}  h(p)W\big( f_p m  \big) \big(P_{\pho,i}{+}a^*(k_if_p m) {+}a(k_if_pm)+\lan f_pm, k_i f_pm \ran   \big) \times \non\\
& &\ph{44444444444}\times (a^*( k_i f_p u)+a( k_i  f_p u))  \phi_p=P_{\pho,i}(-i\pa_{x_i}) \Psi_t(x).  \label{P-pax-eq}
\eeqa
Thus we get from (\ref{P-eq}) and  (\ref{pax-eq})
\beqa
& &(-i\pa_{x_i}- P_{\pho,i}) \Psi_t(x)=  \int d^3p \, e^{i(p\cdot x-E_pt) }  e^{i \thet(p,x,t)}  h(p) W\big( f_p m \big)\times \non\\
& &\ph{444444444444}\times\big(-P_{\pho,i}^{\mrm{w}} +p_i+\pa_{x_i} \thet(p,x,t)+
\mrm{Im}\lan f_pm, f_p\pa_{x_i} m\ran-  \lan f_p,k_i f_p\ran +   2\mrm{Re}\lan f_p, k_i f_p u \ran      \big)   \phi_p\non\\
& &\ph{44444444444444}= \int d^3p \, e^{i(p\cdot x-E_pt) }  e^{i \thet(p,x,t)}  h(p) W\big( f_p m \big) \big(p_i-P_{\pho,i}^{\mrm{w}}) \phi_p,
\label{cancellation}
\eeqa
where we used that
\beqa
\mrm{Im}\lan f_pm, f_p\pa_{x_i} m\ran-  \lan f_p,k_i f_p\ran  + 2\mrm{Re}\lan f_p, k_i f_p u \ran =\mrm{Re}\lan f_p , k_i f_pu\ran=
-\pa_{x_i} \thet(p,x,t).
\eeqa
By iteration of (\ref{cancellation})
\beqa
(-i\pa_{x_i}- P_{\pho,i})^{\ell} \Psi_t(x)= \int d^3p \, e^{i(p\cdot x-E_pt) }  e^{i \thet(p,x,t)}  h(p) W\big( f_p m \big) \big(p_i-P_{\pho,i}^{\mrm{w}} \big)^{\ell} \phi_p. \label{x-P}
\eeqa
Finally, we obtain
\beqa
(a^*(v)+a(v)) \Psi_t(x)=\int d^3p \, e^{i(p\cdot x-E_pt) }  e^{i \thet(p,x,t)}  h(p) W\big( f_p m \big) ( (a^*(v)+a(v))^{\mrm{w}} + 2\mrm{Re}\lan f_pu, v\ran
 ) \phi_p.
\eeqa
One can see, by analogous arguments as in the proof of Lemma~\ref{time-derivative-lemma}~{\bc (a)}, that  all  vectors above are
in $L^2(\real^3_x;\mcF)$: {\bc First, we apply the shift (\ref{shift}) and estimate the term involving $\phi_p-\phi_{p,\si_x}$ with the help
of the bound~(\ref{spectral-bound-intro}). The presence of $H_{\pho}^{\ell}$ in (\ref{spectral-bound-intro}) allows us to control both $P_{\pho,i}$
and the creation and annihilation operators  acting on $\phi_p-\phi_{p,\si_x}$ as for example in the case of (\ref{P-pax-eq}). To the latter
operators we apply the energy bounds (\ref{energy-bounds-eq}) and note that all the resulting $\|\,\cdot \,\|_{\om}$-norms are finite. 
Next, we study the term proportional to $\phi_{p,\si_x}$ using Lemma~\ref{stationary-phase}. Staying with the case of (\ref{P-pax-eq}),
we can rewrite the relevant vector as $P_{\pho,i}(-i\pa_{x_i}\Psi_t^{\si_x})$ and estimate the r.h.s. of (\ref{finite-time-estimate}) 
using Lemmas~\ref{gamma-bounds}, \ref{u-bounds}. In particular, the last part of Lemma~\ref{u-bounds} plays a role here. }
From (\ref{x-P}), (\ref{P-pax-eq}) we also obtain that ${\bc \{ (-i\pa_{x_i})^2\Psi_t(x)\}_{x\in \real^3}}$ is in $L^2(\real^3_x;\mcF)$.
This concludes the proof. \qed\\
\nin\textbf{Proof of Theorem~\ref{lem:time-derivative-final}.} We recall that $\psi_t(x):=\fr{1}{(2\pi)^{3/2}}e^{iHt} e^{{\bc -}iP_{\pho}\cdot x} \Psi_t(x)$. By Lemma~\ref{time-derivative-lemma}, $t\mapsto \Psi_t$
is differentiable in the norm in $L^2(\real^3_x;\mcF)$.  {\rc Next, by applying the Stone
theorem to $e^{iHt}$, we obtain the differentiability of $ t\mapsto \psi_t$ in the norm of $L^2(\real^3_x;\mcF)$, provided that
 the vector  $\{e^{{\bc -}iP_{\pho}\cdot x} \Psi_t(x)\}_{x\in \real^3}\in L^2(\real^3_x;\mcF)$ is in the domain of $H$. This
 is easily checked using Lemma~\ref{Domain-of-H}. In particular,
 to verify that this vector is in the domain of $(-i\nabla_x)^2$, we  apply the Stone theorem to $x\mapsto e^{{\bc-}iP_{\pho}\cdot x}$
 and use  that $\Psi_t$ is in the domain of $P_{\pho}^2$}.   Now we compute
\beqa
\pa_t\psi_t(x)\1&=&\1 \fr{1}{(2\pi)^{3/2}} e^{iHt} iHe^{ {\bc-}iP_{\pho}\cdot x} \Psi_t(x)+\fr{1}{(2\pi)^{3/2}} e^{iHt} e^{{\bc -}iP_{\pho}\cdot x} \pa_t\Psi_t(x)\non\\
\1&=&\1 \fr{1}{(2\pi)^{3/2}}e^{iHt} e^{{\bc-}iP_{\pho}\cdot x} i\bigg( \fr{1}{2}(-i\nabla_x-P_{\pho})^2  \Psi_t(x)+  H_{\pho} \Psi_t(x)+  
 (a^*(v)+a(v))\Psi_t(x)-i\pa_t\Psi_t(x)\bigg)\non\\
\1&=&\1 \fr{1}{(2\pi)^{3/2}}e^{iHt} e^{-iP_{\pho}\cdot x } \int d^3p \,  e^{i (p \cdot x -E_pt  )} e^{i \thet(p,x,t)} i  \ga_{\mrm{int}}(p,x,t)  h(p)
 W(f_p m)	\phi_p,
\eeqa
where in the last step we made use of  the formulas in Lemmas~\ref{time-derivative-lemma}, \ref{Domain-of-H}, 
the fact that $H^{\mrm{w}}_p\phi_p=E_p\phi_p$, and of the relations
\beqa
\lan f_p,|k| f_p\ran-2\mrm{Re}\lan f_p,|k| f_pu\ran+\pa_t\ga(p,x,t)+\mrm{Im}\lan f_pm,f_p\pa_tm\ran=0,
\quad 2\mrm{Re}\lan f_pu(t,x),v\ran= \ga_{\mrm{int}}(p,x,t),
\eeqa
{\rcc where  $v$ appeared in (\ref{H-definition}).}
To show (\ref{pat-bound}), we proceed similarly as in the proof of Lemma~\ref{time-derivative-lemma}:
We choose a $(t,x)$-dependent cut-off as follows: $\si_{(t,x)}=\ka_{\la_0}/(1+|t|+|x|)^M$ where $M\in \nat$ is  fixed. 
We make a shift  $\phi_p=(\phi_p-\phi_{p,\si_{(t,x)}})+\phi_{p,\si_{(t,x)}}$ and insert it into the formula for the norm of $\pa_t\psi_t$:
\beqa
\|\pa_t\psi_t\|_{\hil} \1&\leq &\1 \fr{1}{(2\pi)^{3/2}} \big\| \bigg\{ \int d^3p \,  e^{i (p \cdot x -E_pt  )} e^{i \thet(p,x,t)} i  \ga_{\mrm{int}}(p,x,t)  h(p)  W(f_p (e^{-i|k|  t+ik\cdot x }-1))
	 (\phi_p-\phi_{p,\si_{(t,x)}}) \bigg\}_{\bcc x\in \real^3}  \big\|_{\hil} \non\\
	\1& &\1+ \fr{1}{(2\pi)^{3/2}}\big\| \bigg\{\int d^3p \,  e^{i (p \cdot x -E_pt  )} e^{i \thet(p,x,t)}i  \ga_{\mrm{int}}(p,x,t)  h(p)  W(f_p (e^{-i|k|  t+ik\cdot x }-1))
	 \phi_{p,\si_{(t,x)}} \bigg\}_{\bcc x\in \real^3} \big\|_{\hil}. \label{time-derivative-estimate}
	\eeqa 
We note that by (\ref{spectral-bound-intro}) the term 
involving $(\phi_p-\phi_{p,\si_{(t,x)}}) $ is integrable in $t$ in the norm of $L^2(\real^3_x;\mcF)$ for $M$ sufficiently large. 
Our strategy to estimate the second term on the r.h.s. of (\ref{time-derivative-estimate}) is to  combine Lemma~\ref{stationary-phase}, Lemma~\ref{phase-lemma-zero} and Lemma~\ref{modified-corollary}.
In our case $\uu$ of Lemma~\ref{stationary-phase}  has the form
\beqa
\uu_{(t,x)}(p):=e^{i \thet(p,x,t)} i  \ga_{\mrm{int}}(p,x,t) h(p)W(f_p (e^{-i|k|  t+ik\cdot x }-1))    \phi_{p,\si_{(t,x)}}. \label{u-definition}
\eeqa
We rewrite this expression as follows:
\beqa
\uu_{(t,x)}(p)=    e^{i \thet(p,x,t)} i  \ga_{\mrm{int}}(p,x,t)  h(p)   \hat{\uu}^{\si_{(t,x)}}_{(t,x)}{\rcc (p)}, \quad
 \hat{\uu}^{\si}_{(t,x)}(p):=  W(f_p (e^{-i|k|  t+ik\cdot x }-1) )\phi_{p,\si}.
\eeqa
First, we note that by Lemma~\ref{phase-lemma-zero} below, for $c_0$ as in Lemma~\ref{stationary-phase},
\begin{align}
&|\pa_{p}^{\al}  \ga_{\mrm{int}}(p,x,t)|\leq |\la|^2\fr{c_{\ti M}}{t^{\ti{M}}} \quad \ \  \ph{44}\textrm{ for }\quad |x|/t\leq c_0<1, 
\label{first-estimate-convergence}\\
& |\pa_{p}^{\al}  \ga_{\mrm{int}}(p,x,t)|\leq |\la|^2\fr{c}{t} |\log\,t| \quad \textrm{ for }\quad  |x|/t\geq c_0,
\end{align}
and $|\al|=0,1,2$. Furthermore, we have by Lemma~\ref{gamma-bounds}
\beqa
& &|\pa_p^{\al} e^{i \thet(p,x,t)}|\leq  c (1+\log(1+|t|+|x|))^2.  \label{last-estimate-convergence} 
\eeqa
Given (\ref{first-estimate-convergence})--(\ref{last-estimate-convergence}) and Lemma~\ref{stationary-phase} and Lemma~\ref{u-bounds}, 
{\bc for any $0<\eps<1/2$  we can choose $\la_0$ so small, that}
\beqa
  \|\pa_t\psi_t\|_{\hil}\leq |\la|^2\fr{c}{t^{3/2-\eps}} \label{time-derivative-explicit-bound}
\eeqa
which concludes the proof of (\ref{pat-bound}). Hence, by the Cook method, we obtain the existence 
of the limit $\psi^{+}$.

To see that $\psi^+\neq 0$ under the specified conditions, we write
\beqa
\|\psi^{+,(\la)}\|_{\hil} \geq \|\psi_{t=0}^{(\la)}\|_{\hil}- \int_0^{\infty} dt\, \| \pa_t\psi^{(\la)}_t\|_{\hil}, \label{psi-plus-norm}
\eeqa 
where we included the dependence on $\la$ explicitly in the notation. {\bc We recall that all constants in our discussion
are uniformly bounded in $|\la|\in (0,\la_0]$.}
Thus  by estimate~(\ref{time-derivative-explicit-bound}),
the second term on the r.h.s. of (\ref{psi-plus-norm}) tends to zero as $\la\to 0$. So it suffices to show that $\|\psi_{t=0}^{(\la)}\|_{\hil}$
is bounded from below uniformly in $\la$ from some neighbourhood of zero.  
We collect the relevant ingredients: First, we recall that by \cite[formula (5.2)]{DP12}
\beqa
\|  \phi^{(\la)}_p-\Om\|_{\mcF}\leq c|\la|^{1/4}. \label{vacuum-ground-state}
\eeqa
Furthermore, we obtain from (\ref{f-p-m}), (\ref{f_p-zero-estimate})
\beqa
\|  f^{(\la)}_p (e^{-i|k|t+ik\cdot x }-1)\|_2 \1&\leq&\1 c|\la|(1+\log(1+|t|+|x|))^{1/2},\quad
 |  \thet(p,x,t)| \leq c|\la|^{\bcc 2}(1+\log(1+|t|+|x|)). 
\label{f-p-estimates-one}
\eeqa
Considering the above, we have
 \beqa
 \psi_{t=0}^{(\la)}(x)\1&=&\1\fr{1}{(2\pi)^{3/2}} \int d^3p \,  e^{i p \cdot x   } e^{i \thet^{(\la)}(p,x,0)}  h(p) W(f^{(\la)}_p (e^{ik\cdot x }-1)) \phi^{(\la)}_p \\
\1&=&\1 \fr{1}{(2\pi)^{3/2}}\int d^3p \,  e^{i p \cdot x   } e^{i \thet^{(\la)}(p,x,0)}   h(p) W(f^{(\la)}_p (e^{ik\cdot x }-1)) (\phi^{(\la)}_p-\Om)\\
\1& &\1+\fr{1}{(2\pi)^{3/2}}\int d^3p \,  e^{i p \cdot x   } e^{i \thet^{(\la)}(p,x,0)}   h(p)  \big(W(f^{(\la)}_p (e^{ik\cdot x }-1)) - 1\big) \Om\\
\1& &\1+\fr{1}{(2\pi)^{3/2}}\int d^3p \,  e^{i p \cdot x   } \big(e^{i \thet^{(\la)}(p,x,0)}-1\big)   h(p)  \Om \\
\1& &\1 +\fr{1}{(2\pi)^{3/2}}\int d^3p \,  e^{i p \cdot x   }  h(p)  \Om.
\eeqa
Thus it is manifest from estimates (\ref{f-p-estimates-one}), (\ref{vacuum-ground-state}), {\rc combined with an argument as in (\ref{A-9})}, that
\beqa
\psi_{t=0}^{(\la)}(x)=({\bc F^{-1}} h)(x)\Om+O(|\la|^{1/4} (1+\log(1+|x|)) ),
\eeqa
where $\| O(|\la|^{1/4} (1+\log(1+|x|)) )\|_{\mcF}\leq c|\la|^{1/4}  (1+\log(1+|x|))$. Clearly, we can write for 
any compact subset $\De\subset \real^3$
\beqa
\|\psi_{t=0}^{(\la)}\|_{\hil} \1&\geq&\1 \bigg( \int_{\De} d^3x\, \|\psi_{t=0}^{(\la)}(x)\|^2_{\mcF} \bigg)^{1/2}\non\\
\1 &\geq& \1  \bigg( \int_{\De} d^3x \,|(F^{-1} h)(x)|^2  \bigg)^{1/2}-  c|\la|^{1/4} \bigg( \int_{\De} d^3x\, (1+\log(1+|x|))^2  \bigg)^{1/2}.
\label{small-lambda}
\eeqa
For any  $\De$ intersecting with the support of $F^{-1} h$ {\bcc the first term in the second line of (\ref{small-lambda}) is positive and independent
of $\la$. As the second term tends to zero as $\la\to 0$, this concludes the proof.} \qed
\bel\label{phase-lemma-zero}  
Consider the expression 
\beqa
\ga_{\mrm{int}}(p,x,t):= 2 \int d^3k\, f_p(k)^2(|k|-k\cdot \nabla E_p) \cos(|k|t-k\cdot x). \label{ti-ga-formula}
\eeqa
The following bounds hold:

\begin{enumerate}

\item[(a)] Fix some $0<c_0<1$. For any $M\in \nat$ there exists a constant $c_M$, uniform in $p\in S$,  s.t. 
\beqa
\sup_{(|x|/t)\leq c_0 }|  \ga_{\mrm{int}}(p,x,t)   | \leq |\la|^2\fr{c_M}{t^M}. 
\eeqa
The same is true if the supremum is taken over $|x|/t\geq c_1>1$. 

\item[(b)] For all $p\in S$ and $(t,x)\in \real^4$
 \beqa
 |   \ga_{\mrm{int}}(p,x,t)     |\leq |\la|^2 \fr{c}{t} |\log\,t|.   
 \eeqa
\end{enumerate}
Analogous estimates hold if we replace $p\mapsto f_p(k)^2(|k|-k\cdot \nabla E_p)$  in (\ref{ti-ga-formula})
by its arbitrary derivatives w.r.t. $p$.
\eel
\proof Proceeding to spherical coordinates, we have
\beqa
\ga_{\mrm{int}}(p,x,t)= \int d\Om(e_k)  \int_0^{\infty} d|k| \, f(|k|, e_k,p)    \cos(|k|t(1- e_k\cdot \mrm{v} )), \ \ 
f(|k|, e_k,p):=|\la|^2\fr{\chi_{\ka}(k)^2}{ 2}\fr{1}{(1-e_k\cdot \nabla E_{p} )},
\eeqa
where we set $\mrm{v}:=x/t$. We suppose  that  $||\mrm{v}|-1|\geq \eps> 0$ and consider part (a) of the lemma. By integrating by parts w.r.t. $|k|$ and exploiting that sine vanishes at zero, we obtain
\beqa
 \ga_{\mrm{int}}(p,x,t)=  -\int d\Om(e_k)  \int_0^{\infty} d|k| \, \pa_{|k|}  f(|k|, e_k,p)  \fr{1}{t(1-  e_k\cdot \mrm{v}) }  \sin(|k|t(1- e_k\cdot \mrm{v} )).
\eeqa
Concerning (a), we can continue integrating by parts, exploiting that $\pa_{|k|}f$ vanishes in a fixed neighbourhood of zero. The fact that $(1-  e_k\cdot \mrm{v})$
is never zero in this case gives the claim.

Proceeding to (b), we suppose that $||\mrm{v}|-1|\leq \eps$ for some $0<\eps<1$, in particular $|\mrm{v}|$ is isolated from zero by some interval independent of $p,x,t$. We choose the third axis in the direction of $\mrm{v}$ and write
\beqa
& &\!\!\!\!\!\!\ga_{\mrm{int}}(p,x,t)\non\\
\1&=&\1 \int_{|k|\geq 1/t} d|k| \int_0^{2\pi} {\rcc d\vp} \int_{-1}^{1} d\cos(\theta)  \, f(|k|, e(\cos(\theta), {\rcc \vp}),p)    \cos(|k|t(1-  |\mrm{v}| \cos(\theta) ))+O(t^{-1})\non\\
\1&=&\1 -\int_{|k|\geq 1/t} d|k| \int_0^{2\pi} {\rcc d\vp} \int_{-1}^{1} d\cos(\theta)  \, f(|k|, e(\cos(\theta), {\rcc \vp}),p) \fr{1}{t |k| |\mrm{v}|}\fr{d}{d\cos(\theta)}   \sin(|k|t(1-  |\mrm{v}| \cos(\theta) ))+O(t^{-1})\non\\
\1&=&\1 -\int_{|k|\geq 1/t} d|k| \int_0^{2\pi} {\rcc d\vp}   \, f(|k|, e(\cos(\theta), {\rcc \vp}),p) \fr{1}{t|k| |\mrm{v}|} \sin(|k|t(1-  |\mrm{v}| \cos(\theta) ))|^{\cos\theta=1}_{\cos\theta=-1 }+O(t^{-1})\non\\
\1& &\1 +\int_{|k|\geq 1/t} d|k| \int_0^{2\pi} {\rcc d\vp} \int_{-1}^{1} d\cos(\theta)  \,\bigg( \fr{d}{d\cos(\theta)} f(|k|, e(\cos(\theta), {\rcc \vp}),p)\bigg) \fr{1}{ t|k|  |\mrm{v}|}  \sin(|k|t(1-  |\mrm{v}| \cos(\theta) )).
\eeqa
By estimating $|\sin(|k|t(1-  |\mrm{v}| \cos(\theta) )) |\leq 1$ everywhere above and using that the integration in $|k|$ is over a compact set, the claim follows from
\beqa
\fr{1}{t}\int_{\kappa \geq |k|\geq 1/t} d|k| \fr{1}{|k|}\leq \fr{c'}{t}|\log(t)|.
\eeqa
This concludes the proof. \qed

\section{Conclusions}
\setcounter{equation}{0}

In this paper we proposed a new construction of infraparticle states in the massless Nelson model.
The approximating sequence does not involve infrared cut-offs and the proof of convergence is relatively
simple:  Taking  the spectral results from \cite{AH12,DP12, DP16} for granted, it amounts to the Cook method
combined with  the stationary phase method, like for basic Schr\"odinger operators. It is legitimate to ask how
the new infraparticle state compares with the established knowledge on the infrared problem in the Nelson model.
To partially answer this question we provide some heuristic remarks on the relation of our states to the 
Faddeev-Kulish approach.  First, we note that the asymptotically dominant part of the wave packet (\ref{eq:infraparticle})
should propagate along the ballistic trajectory $x=\nabla E_p t$, thus $\psi_t$ should have the same limit as
\beqa
\psi^{\mrm{D}}_t(x):=e^{iHt}\fr{1}{(2\pi)^{3/2}}\int d^3p\, h(p)\, e^{-i(E_p +H_{\pho} )t }  e^{i \thet(p,\nabla E_pt,t)}   
 W\big( f_p (1 - e^{i|k|t-ik\cdot \nabla E_p t}) \big) e^{iH_{\pho}t } \fr{1}{(2\pi)^{3/2}}e^{i(p- P_{\pho})\cdot x} \phi_p. \label{Dollard-approach}
 \eeqa
To proceed, let us second quantize also the electrons, denote their creation and annihilation 
operators by $b^{(*)}$ and the common vacuum of the electrons and photons by $\Om$. Expressing $\phi_p\in \mcF$ by its
$n$-particle wave functions $f^{n}_{\mrm{w},p}$ we  define its renormalized creation operator in a standard manner \cite{Al73}:  
\beqa
\hat{b}^*_{\mrm{w}}(p):=\sum_{\m=0}^{\infty}\fr{1}{ \sqrt{\m!} }\int d^{3\m}k\, f^{\m}_{\mrm{w},p}(k_1,\ldots, k_{\m})a^*(k_1)\ldots a^*(k_{\m}) \, b^*(p- (k_1+\cdots+k_{\m})),
\eeqa
so that $ \fr{1}{(2\pi)^{3/2}} e^{i(p- P_{\pho})\cdot x} \phi_p$ can be identified with $\hat{b}^*_{\mrm{w}}(p)\Om$. Now recalling that $f_p(k)=v(k)\fr{1}{|k|-k\cdot \nabla E_p}$, we can write
\beqa
W\big(f_p(1 - e^{i|k|  t-ik\cdot \nabla E_p t}) \big)\1&=&\1  
\exp{\bigg(-i  \int_0^td \tau\,   e^{iH_{\pho}\tau }\big\{ a^*\big( v  e^{  -ik\cdot \nabla E_p \tau}  \big) +   a\big( v e^{ -ik\cdot \nabla E_p \tau}  \big)   \big\} e^{-iH_{\pho}\tau }  \bigg) } \non\\
\1&=&\1e^{iC_pt} e^{-i\ga(p,\nabla E_pt,t)}  \mrm{Texp}\bigg({-i  \int_0^td \tau\,   e^{iH_{\pho}\tau }\big\{ a^*\big( v  e^{  -ik\cdot \nabla E_p \tau}  \big) +   a\big( v e^{ -ik\cdot \nabla E_p \tau}  \big)   \big\} e^{-iH_{\pho}\tau }   \bigg)},
\eeqa
where $C_p:=\int d^3k\,\fr{v(k)^2}{|k|-k\cdot \nabla E_p}$ is finite and the time-ordered exponential $U_{\mrm{D}}(t):=\mrm{Texp}(\,\ldots\,)$ is the Dollard modifier 
of the Nelson model, cf \cite[formula (3.6)]{Dy17}. Thus (\ref{Dollard-approach}) can be rewritten as  
\beqa
\psi^{\mrm{D}}_t=e^{iHt}\int d^3p\, h(p)\, e^{-i(E_p +H_{\pho}-C_p)t }     
  U_{\mrm{D}}(t)   e^{iH_{\pho}t }\hat{b}_{\mrm{w}}^*(p)\Om. \label{FK-Nelson}
\eeqa
We recall from \cite{Dy17}, that a direct application of the Faddeev-Kulish prescription to the Nelson model leads
to a formula which differs from (\ref{FK-Nelson}) only by  {\bcc a} substitution $\hat{b}_{\mrm{w}}^*(p)\to b^*(p)$.
We believe that this discrepancy can be attributed to the quantum mechanical origin of the Dollard formalism which
makes it difficult to reconcile with the electron mass renormalization present in the model. We think that 
formula~(\ref{FK-Nelson}) is a correct implementation of the Faddeev-Kulish formalism in the Nelson model
and hope that the findings of the present paper will lead to a rigorous proof of  convergence {\bc of $\psi^{\mrm{D}}_t$} as $t\to \infty$.

\appendix
\section{Energy bounds and derivatives of the Weyl operators} \label{Weyl-differentiability}
\setcounter{equation}{0}

We introduce the following subspace of $L^2(\real_k^3)$:
\beqa
L^2_{\om}(\real_k^3):=\{  f\in L^2(\real_k^3)\,|\, \|f\|_{\om}:=\|(1+|k|^{-1/2})f\|_2<\infty  \}. \label{L-2-omega}
\eeqa
{\mg We recall that  $N:=\dGa(1), H_{\pho}:=\dGa(|k|)$ and state the standard energy and number bounds \cite{BR2}:}
\bel\label{energy-bounds-lemma} Let $f_1, \ldots, f_{{\bc n}}\in  L^2(\real_k^3)$. Then
\beqa
\|a^{(*)}(f_1)\ldots a^{(*)}(f_{{\bc n}})(1+N )^{-n/2}\|_{\mcF}\leq c_n \|f_1\|_{2} \ldots \|f_n\|_{2}. \label{number-bounds-eq}
\eeqa
Let $f_1, \ldots, f_{{\bc n}}\in  L^2_{\om}(\real_k^3)$. Then
\beqa
\|a^{(*)}(f_1)\ldots a^{(*)}(f_{{\bc n}})(1+H_{\pho})^{-n/2}\|_{\mcF}\leq c_n \|f_1\|_{\om} \ldots \|f_n\|_{\om}. \label{energy-bounds-eq}
\eeqa
\eel
{\bcc \nin Formula~(\ref{W-der-zero}) below is also well-known, but we provide a proof for the reader's convenience.}
\bel\label{W-der-lemma} Let $\real\ni s\mapsto F_s\in L^2_{\om}(\real_k^3)$ be  differentiable in the norm $\|\,\cdot\,\|_{\om}$.
Then $s\mapsto W(F_s)\psi$,  $\psi\in D(H_{\pho}^{1/2})$,  is differentiable in the norm of $\mcF$  
and
\beqa
\pa_sW(F_s)\psi= W(F_s)(a^*(\pa_sF_s)-a(\pa_s F_s)+i\mrm{Im}\lan F_s, {\bc \pa_s}F_s\ran)\psi. \label{W-der-zero}
\eeqa 
Also, $s\mapsto a^{(*)}(F_s)\psi$ {is} differentiable w.r.t. $s$ in the norm of $\mcF$ and $\pa_s a^{(*)}(F_s)\psi=a^{(*)}(\pa_sF_s)\psi.$ 
If $\psi\in D(N^{1/2})$ then {\bc  analogous statements hold} for $s\mapsto F_s$ differentiable in the norm $\|\,\cdot\,\|_{2}$.
\eel
\proof Using the Weyl relations $W(F)W(G)=e^{-i\mrm{Im}\lan F,G\ran }W(F+G)$, $F,G\in L^2(\real^3_k)$,
\beqa
\fr{1}{\De s} \big( W(F_{s+\De s})-W(F_s) \big)\1&=&\1W(F_s)\fr{1}{\De s}(W(-F_s)W(F_{s+\De s})-1)\non\\
\1&=&\1 W(F_s)\fr{1}{\De s}( e^{i\mrm{Im}\lan F_s, F_{s+\De s}-F_s\ran }W(     F_{s+\De s} -F_s  )-1)\non\\
\1&=&\1 W(F_s)\fr{1}{\De s}\big(e^{i\mrm{Im}\lan F_s, F_{s+\De s}-F_s\ran } -1\big) W(  F_{s+\De s} -F_s  )\label{W-der-one} \\
\1& &\1+ W(F_s) \fr{1}{\De s}\big(  W(     F_{s+\De s} -F_s  )-1\big). \label{W-der-two}
\eeqa
Considering (\ref{W-der-one}), we obtain immediately
\beqa
\lim_{\De s\to 0}\fr{1}{\De s}\big(e^{i\mrm{Im}\lan F_s, F_{s+\De s}-F_s\ran } -1\big)= i\mrm{Im}\lan F_s, \pa_sF_s\ran. \label{imaginary}
\eeqa
Furthermore, it is easy to see that in the norm of $\mcF$
\beqa
\lim_{\De s\to 0}W(F_{s+\De s} -F_s)\psi=\psi. \label{W-diff-property}
\eeqa
In fact, denoting $\Phi(F):=a^*(-iF)+a(-iF)$, we can write
\beqa
W(F_{s+\De s} -F_s)\psi=\psi+ \bigg\{\fr{e^{i\Phi(F_{s+\De s} -F_s)} -1}{\Phi(F_{s+\De s} -F_s)} \bigg\}\Phi(F_{s+\De s} -F_s) \psi. \label{A-9}
\eeqa
By the spectral theorem, the norm of the expression in curly bracket above is bounded uniformly in $\De s$. On the other
hand,  by the assumed form of differentiability
\beqa
\Phi(F_{s+\De s} -F_s) \psi=\De s\, \Phi(  \pa_sF_s)\psi+\Phi( {\mg o(\Delta s)} )\psi, \label{Phi-bound}
\eeqa
where $\pa_sF_s\in L^2_{\om}(\real^3_k)$ and the rest term satisfies 
\beqa
\lim_{\De s\to 0} \fr{ \| {\mg  o(\De s)}\|_{\om}}{\De s}=0. \label{o(s)}
\eeqa
Thus, by the energy bounds of Lemma~\ref{energy-bounds-lemma} we obtain that (\ref{Phi-bound}) tends to zero in {\bc the norm of $\mcF$}
as $\De s\to 0$ which gives (\ref{W-diff-property}). 

Concerning (\ref{W-der-two}), we write again $F_{s+\De s} -F_s=\De s \, \pa_sF_s+{\mg o( \De s)}$, {\bc which gives}
\beqa
\fr{1}{\De s}\big(  W(     F_{s+\De s} -F_s  )-1\big)\psi=\fr{1}{\De s}\big(  e^{-i\mrm{Im}\lan\De s\, \pa_sF_s, {\mg o(\De  s)}\ran} W( \De s \pa_sF_s)W({\mg o(\De s)}) -1\big){\bc \psi}.
\eeqa
{\bc To take the limit $\De s\to 0$ above, we note}
\beqa
\lim_{\De s\to 0}\fr{1}{\De s} \big(e^{-i\mrm{Im}\lan\De s \pa_sF_s, {\mg o(\De s)}\ran}-1)=0, \quad  
\lim_{\De s\to 0}\fr{1}{\De s}\big( W({\mg o(\De s)})-1\big)\psi=0,
\eeqa
where the latter limit is computed as in (\ref{Phi-bound}) using (\ref{o(s)}).  {\bc Also,  we exploit  that by the Stone theorem}
\beqa
\lim_{\De s\to 0} \fr{1}{\De s}(W( \De s \pa_sF_s)-1)\psi=i\Phi(\pa_sF_s)\psi.  \label{field}
\eeqa
{Finally}, substituting (\ref{field}), (\ref{imaginary}) to (\ref{W-der-one}), (\ref{W-der-two}), we obtain (\ref{W-der-zero}). 
The last statement of the lemma is proven by analogous  arguments. \qed

\section{Proof of Lemma~\ref{modified-corollary}} \label{first-appendix}
\setcounter{equation}{0}

We write $\phi_{p,\si}=\{ f^{\m}_{\mrm{w},p,\si} \}_{\m\in \nat_0}$ in terms of the Fock space wave functions.
Given Lemma~\ref{wavefunctions-lemma} and formula~(\ref{g-function})  below, we can write
\beqa
\| H_{\pho}^{\ell_1} N^{\ell_2} \pa_{p}^{\al}\phi_{p,\si}\|_{\mcF} \leq  \bigg(\sum_{\m=0}^{\infty} \m^{2(\ell_1+\ell_2)} \|\pa_{p}^{\al} 
f^{\m}_{\mrm{w}, p,\si}\|_2^2\bigg)^{1/2}
 \leq \fr{c}{\si^{\de'_{\la_0}}} \bigg(\sum_{\m=0}^{\infty} \fr{1}{\m!}  (c_{\ell_1,\ell_2}\la)^{\m}|\log\,\si|^{\m}   \bigg)^{1/2}\leq \fr{c}{\si^{\de_{\la_0} } }, \label{N-computation}
\eeqa
for some constants $c_{\ell_1,\ell_2}$ and $\de_{\la_0}>0$ which tends to zero as $\la_0\to 0$. {\rcc To handle the powers of $H_{\pho}$ we used that the UV cut-off $\ka=1$ and consequently the wave functions $f^{\m}_{\mrm{w},p,\si}$ are supported in unit balls in each variable $k_1,\ldots, k_n$ separately.} This gives Lemma~\ref{modified-corollary}.

In preparation for the proof of Lemma~\ref{wavefunctions-lemma}, we state a general relation for wave functions of a Fock space vector: 
\beqa
f^{\m}_{\mrm{w},p,\si}(k_1,\ldots,k_{\m})=\fr{1}{\sqrt{\m!}}\lan \Om,a(k_1)\ldots a(k_{\m})\phi_{p,\si}\ran. \label{wave-function-def}
\eeqa
This formula is meaningful by considerations in  \cite[Appendix D]{DP12}.
Let us now introduce the following auxiliary functions:
\beqa
& &g_{\si}^{0}:=c \quad \textrm{ and }\quad   g_{\si}^{\m}(k_1,\ldots, k_{\m}):=\prod_{i=1}^{\m}\fr{ c\g  \chi_{[\si,\kas)}(k_i)  }{|k_i|^{3/2}}, \m\geq 1, \label{g-function}  
\eeqa
where $\kas:=(1-\eps_0)^{-1}\ka$ is slightly larger than $\kappa$ {\mg and $0<\eps_0<1$ was introduced below (\ref{H-definition})}. 
\bel\label{wavefunctions-lemma} The following estimates hold
\beqa
|\pa_{p}^{\al}f^{\m}_{\mrm{w},p,\si}(k_1,\ldots,k_{\m})|\leq \fr{1}{\sqrt{\m!}} \bigg( \fr{1}{\si^{\de_{\la_0} } }\bigg)^{|\al|} 
g_{\si}^{\m}(k_1,\ldots, k_{\m}) \quad\textrm{for} \quad |\al|\leq 2. \label{wave-function-lemma}
\eeqa
\eel
\proof {\bcc In \cite[formula (4.42)]{DP16}  the following functions are introduced}\footnote{ 
{\rcc For consistency with the notation from \cite{DP12,DP16, DP19}, we use similar symbols for several different functions: $f_{p,\si}$ are defined in (\ref{Bogolubov-cut}), $\hat f^{\m}_{p,\si}$ in (\ref{hat-f-p-sigma}) and  $f^{\m}_{\mrm{w},p,\si}$ are the wave functions of $\phi_{p,\si}$.} }
\beqa
\hat f^{\m}_{p,\si}(k_1, \ldots, k_{\m}):= W^*(f_{p,\si}) a(k_1)\ldots a(k_{\m})\phi_{p,\si}, \label{hat-f-p-sigma}
\eeqa 
where $W(f_{p,\si}):=e^{a^*(f_{p,\si})-a(f_{p,\si}) }$ {\mg and the r.h.s. above is well-defined by considerations from \cite[Appendix D]{DP12}}.  In Proposition~4.7 of \cite{DP16} and in the 
subsequent discussion in this reference the following bounds are shown
\beqa
\| \pa^{\al}_p\hat f^{\m}_{p,\si}(k_1,\ldots, k_{\m})\|_{\mcF}\leq \bigg( \fr{1}{\si^{\de_{\la_0} } }\bigg)^{|\al|}  g^{\m}_{\si}(k_1,\ldots, k_{\m})\quad \textrm{for} \quad |\al|\leq 2. \label{borrowed-estimate}
\eeqa
{\mg In view of (\ref{g-function}) the r.h.s. depends on numerical constants, whose dependence on various parameters is specified  at the end of  Section~\ref{Preliminaries}.} We note that for $|\al|=0$ (\ref{wave-function-lemma})  follows immediately from (\ref{borrowed-estimate}) 
{\mg and (\ref{wave-function-def}).}

As for the case $|\al|=1$, we can write
\beqa
\pa_{p_j}  (a(k_1)\ldots a(k_{\m})  \phi_{p,\si})=\pa_{p_j}(W(f_{p,\si}) \hat f^{\m}_{p,\si}(k_1, \ldots, k_{\m})). 
\eeqa
The term in which the derivative acts on $\hat f^{\m}_{p,\si}(k_1, \ldots, k_{\m})$ is immediately estimated using (\ref{borrowed-estimate})
{\mg for $|\al|=1$}. As for the remaining term, we estimate
\beqa
& &\| \pa_{p_j}(W(f_{p,\si})) \hat f^{\m}_{p,\si}(k_1, \ldots, k_{\m})\|_{\mcF} \non\\
& &\ph{44444444444}  \leq  2\| a(\pa_{p_j} f_{p,\si}) {\bcc W(f_{p,\si})} \hat f^{\m}_{p,\si}(k_1, \ldots, k_{\m})\|_{\mcF}+
\| \pa_{p_j} f_{p,\si}\|_2 \,  \| \hat f^{\m}_{p,\si}(k_1, \ldots, k_{\m})\|_{\mcF}\label{derivative-W} \\
& &\ph{44444444444}  \leq  2\int d^3k_0\, |(\pa_{p_j} f_{p,\si})(k_0)| \, \|\hat f^{\m+1}_{p,\si}(k_0,k_1, \ldots, k_{\m})\|_{\mcF}+{\mg c}|\log(\si)|^{1/2} \,\| \hat f^{\m}_{p,\si}(k_1, \ldots, k_{\m})\|_{\mcF}.\quad\quad\quad
\label{derivative-W-one}
\eeqa
For differentiability of the Weyl operator we refer to Lemma~\ref{W-der-lemma} and  the fact that  
 $\hat f^{\m}_{p,\si}$ is in the domain of $H_{\pho}^{1/2}$ (cf.  \cite[formula (D.8)]{DP12}). {\mg The bound on $\| \pa_{p_j} f_{p,\si}\|_2$
 follows from   Lemma~\ref{new-lemma}. }
This, together with (\ref{borrowed-estimate}), gives (\ref{wave-function-lemma}) for $|\al|=1$. 

Now  we consider the case $|\al|=2$. Again, we can write
\beqa
\pa_{p_j}\pa_{p_i}  \big(a(k_1)\ldots a(k_{\m})  \phi_{p,\si}\big)= \pa_{p_j}\pa_{p_i} \big( W(f_{p,\si})   \hat f^{\m}_{p,\si}(k_1, \ldots, k_{\m})\big).\label{second-der-est}
\eeqa
The term in which both derivatives act on $\hat f^{\m}_{p,\si}$ is immediately estimated using (\ref{borrowed-estimate}). Let us 
consider the term in which one derivative acts on $W(f_{p,\si})$ and another on $\hat f^{\m}_{p,\si}$. {\bcc Similarly} as in (\ref{derivative-W}),
we have
\beqa
\|\pa_{p_j} \big( W(f_{p,\si})\big)  \pa_{p_i}\hat f^{\m}_{p,\si}(k_1, \ldots, k_{\m})\|_{\mcF}\leq  2\| a(\pa_{p_j} f_{p,\si}) \pa_{p_i} \hat f^{\m}_{p,\si}(k_1, \ldots, k_{\m})\|_{\mcF}+
\| \pa_{p_j} f_{p,\si}\|_2 \,  \| \pa_{p_i} \hat f^{\m}_{p,\si}(k_1, \ldots, k_{\m})\|_{\mcF}. \label{W-almost-last-term}
\eeqa
The last term on the r.h.s. of (\ref{W-almost-last-term}) clearly satisfies the required bound {\mg by (\ref{borrowed-estimate}) and Lemma~\ref{new-lemma}}.  As for the first term above, we note
that
\beqa
a(\pa_{p_j} f_{p,\si})\pa_{p_i} \hat f^{\m}_{p,\si}(k_1, \ldots, k_{\m})  
\1&=&\1- \bigg( \int d^3k_0\, (\pa_{{\mg p_j}} f_{p,\si})(k_0) f_{p,\si}(k_0)\bigg) \, \pa_{p_i} (W(f_{p,\si})^* a(k_1)\ldots a(k_{\m})\phi_{p,\si})\non\\
\1& &\1+\int d^3k_0\, (\pa_{{\mg p_j}} f_{p,\si})(k_0)\,  \pa_{p_i} (W(f_{p,\si})^* a(k_0)a(k_1)\ldots a(k_{\m})\phi_{p,\si}),
\eeqa
where we first computed the derivative {\mg of} $\hat f^{{\mg n}}_{p,\si}$ and  then used $a(g) W(f_{p,\si})^*=W(f_{p,\si})^*(a(g)-\lan g, f_{p,\si}\ran)$ for $g={\mg \pa_{p_j}f_{p,\si}}$. 
The last expression is immediately estimated using (\ref{borrowed-estimate})
for $|\al|=1$.  We still have to estimate a contribution to (\ref{second-der-est}), where both derivatives act on $W(f_{p,\si})$:
\beqa
& &\| \pa_{p_i}\pa_{p_j}(W(f_{p,\si})) \hat f^{\m}_{p,\si}(k_1, \ldots, k_{\m})  \|_{\mcF} \non\\
& &\ph{4444444444444} \leq  \| (a^*(\pa_{p_j}f_{p,\si} )-a(\pa_{p_j}f_{p,\si} ))(a^*(\pa_{p_i}f_{p,\si} )-a(\pa_{p_i}f_{p,\si} )) 
W(f_{p,\si})\hat f^{\m}_{p,\si}(k_1, \ldots, k_{\m})  \|_{\mcF}\,\,\,\, \\
& & \ph{44444444444444}{\mg+\| (a^*(\pa_{p_i}\pa_{p_j}f_{p,\si} )-a( \pa_{p_i}\pa_{p_j}f_{p,\si} )) 
W(f_{p,\si})\hat f^{\m}_{p,\si}(k_1, \ldots, k_{\m})  \|_{\mcF} }.
\eeqa
This expression can be estimated by a linear combination {\mg of} terms of the form:
\beqa
& &\|\pa_{p}^{\mg \al_1}f_{p,\si}\|_2 \,\|\pa_{p}^{\mg \al_2}f_{p,\si}\|_2 \, \| \hat f^{\m}_{p,\si}(k_1, \ldots, k_{\m})  \|_{\mcF},  \label{first-term-of-three} \\
& &\|\pa_{p}^{\mg \al_1}f_{p,\si}\|_2\,\|a(\pa_{p}^{\mg \al_2} f_{p,\si}) W(f_{p,\si}) \hat f^{\m}_{p,\si}(k_1, \ldots, k_{\m})\|_{\mcF}, \label{second-term-of-three}\\
& &\| a( \pa_{p}^{\mg \al_1}f_{p,\si})a( \pa_{p}^{\mg \al_2}f_{p,\si} ) W(f_{p,\si})  \hat f^{\m}_{p,\si}(k_1, \ldots, k_{\m}) \|_{\mcF}, \label{third-term-of-three}
\eeqa
{\mg where $|\al_1|, |\al_2|\leq 2$}.
Expression (\ref{first-term-of-three}) is estimated using (\ref{borrowed-estimate}) for $|\al|=0$ {\mg and Lemma~\ref{new-lemma}}. 
Expression (\ref{second-term-of-three}) is estimated as in (\ref{derivative-W-one}). As for (\ref{third-term-of-three}),
it can be bounded by
\beqa
(\ref{third-term-of-three})\leq \int d^3k' d^3k''\, |\pa_{p}^{\mg \al_1}f_{p,\si}(k') | \,  |\pa_{p}^{\mg \al_2}f_{p,\si}(k'') | \, \|  \hat f^{\m+2}_{p,\si}(k', k'', k_1, \ldots, k_{\m}) \|_{\mcF},
\eeqa
which is estimated with the help of (\ref{borrowed-estimate}) for $|\al|=0$ and {\mg Lemma~\ref{new-lemma}}. \qed
\section{Proof of estimate (\ref{spectral-bound-intro})} \label{approximation-appendix}
\setcounter{equation}{0}

\bel\label{Hamiltonian-lemma} For any $\ell\in \nat_0$ the maximal coupling constant $\la_0>0$ {\mg can be chosen sufficiently small, 
so that  there exists a constant $c$ such that}
\beqa
 \| H_{\pho}^{\ell}(\phi_p-\phi_{p,\si})\|_{\mcF} \leq {\mg c}\si^{1/5}. \label{shift-Hamiltonian-two-zero}
\eeqa
\eel
\proof First, we will prove
 \beqa
\|(H_{p}^{\mrm{w}})^{\ell}(\phi_p-\phi_{p,\si})\|_{\mcF} \leq {\mg c}\si^{1/5}. \label{shift-Hamiltonian-two}
\eeqa
To this end, we recall from \cite{Pi03}, \cite[Lemma 3.6]{DP12} the form of the modified Hamiltonian on 
$D(P_{\pho}^2+H_{\pho})$
\beqa
H_{p,\si}^{\mrm{w}}=\h \Ga_{p,\si}^2+\int d^3k\, \al_{p,\si}(e_k)|k| a^*(k)a(k)+c_p^{\si}, \label{infrared-Hamiltonian}
\eeqa
where
\beqa
& &\Ga_{p,\si}:=\nabla E_{p,\si}-(p-P_{\mrm{f}, \si}^{\mrm{w}}),\quad P_{\mrm{f}, \si}^{\mrm{w}}
:=W(f_{p,\si}) P_{\mrm{f}} W(f_{p,\si})^*, \quad
\al_{p,\si}(e_k):=(1-e_{k}\cdot \nabla E_{p,\si}), \\
& &c_p^{\si}:=\h p^2-\h(p-\nabla E_{p,\si})^2-\la^2\int d^3k \fr{\chi_{{\mg [\si,\ka)}}(k)}{2|k|^2\al_{p,\si}(e_k) }.
\eeqa
The corresponding quantities at $\si=0$ are denoted by dropping $\si$ in the notation. We will use
the standard bounds from \cite{Pi03} 
\beqa
|E_p-E_{p,\si}|\leq c\si, \quad |\nabla E_p-\nabla E_{p,\si}|\leq c\si^{1/4}, \quad \|\phi_p-\phi_{p,\si}\|_{\mcF}\leq c\si^{1/2}, \label{standard-bounds}
\eeqa
see also \cite[Theorem 2.1 (b), Corollary 5.6]{DP12}, \cite[Proposition A.2]{DP19}.

Now we proceed by induction: for $\ell=0$ the estimate holds by the third bound in (\ref{standard-bounds}). For
the inductive step we compute
\beqa
\| (H_{p}^{\mrm{w}})^{\ell} (\phi_p-\phi_{p,\si})\|_{\mcF}\1&=&\1
 \|(H_{p}^{\mrm{w}})^{\ell-1} ( E_{p}\phi_p-  H^{\mrm{w}}_{p} \phi_{p,\si})\|_{\mcF}\non\\
\1&\leq &\1\|(H_{p}^{\mrm{w}})^{\ell-1}(E_p\phi_p-E_{p,\si}\phi_{p,\si})\|_{\mcF}+
\| (H_{p}^{\mrm{w}})^{\ell-1} (H_{p}^{\mrm{w}} - H_{p,\si}^{\mrm{w}})  \phi_{p,\si}\|_{\mcF}. 
\label{Hamiltonian-shift}
\eeqa
The first term on the r.h.s. of (\ref{Hamiltonian-shift}) is $O(\si^{1/5})$ by the induction hypothesis and the first estimate in (\ref{standard-bounds}). Concerning the last term on the r.h.s. of (\ref{Hamiltonian-shift}), we note that there are three contributions to $H_{p}^{\mrm{w}} - H_{p,\si}^{\mrm{w}}$ coming from the three terms in the Hamiltonian  (\ref{infrared-Hamiltonian}). They have the following properties: First,  by (\ref{standard-bounds}), $|c_p-c_p^{\si}|\leq c\si^{1/4}$.
Thus, by Lemma~\ref{dGa-lemma} {\mg below}, 
\beqa
\| (H_{p}^{\mrm{w}})^{\ell-1} (c_p-c_p^{\si})\phi_{p,\si}\|_{\mcF}\leq c\si^{1/4-\de_{\la_0}}.
\eeqa
{\mg Clearly, for $\la_0>0$ sufficiently small, the last expression is $O(\si^{1/5})$.}
The second contribution is $\dGa((\al_{p,\si}(e_k)-\al_{p}(e_k)) |k| )$, where $|\al_{p,\si}(e_k)-\al_{p}(e_k)|\leq c\si^{1/4}$ by (\ref{standard-bounds}). Thus Lemma~\ref{dGa-lemma} gives
\beqa
\| (H_{p}^{\mrm{w}})^{\ell-1}  \dGa((\al_{p,\si}(e_k)-\al_{p}(e_k)) |k| )  \phi_{p,\si}\|_{\mcF}\leq c\si^{1/4-\de_{\la_0}}.
\eeqa
Concerning the  third contribution,  $(\Ga_{p}^2-\Ga_{p,\si}^2)$,
we note that, on $D(P_{\pho}^2+H_{\pho})$,
\beqa
(P_{\mrm{f}, \si}^{\mrm{w}})_i \1&=&\1 W(f_{p,\si}) P_{\mrm{f},i} W(f_{p,\si})^*={\crr (P_{\mrm{f}})_i} -a^*(k_i f_{p,\si})-a( k_i f_{p,\si})
+ \lan f_{p,\si}, k_i f_{p,\si}\ran   \non\\
\1&=&\1 {\crr (P_{\mrm{f}}^{\mrm{w}})_{i}}-a^*(k_i (f_{p,\si}-f_p))-a( k_i (f_{p,\si}-f_p) ) +( \lan f_{p,\si}, k_i f_{p,\si}\ran  -  \lan f_{p}, k_i f_{p}\ran).    
\eeqa
Consequently,  $\Ga_{p,\si}=\Ga_p+\De\Ga_{p,\si}$, where
\beqa
(\De\Ga_{p,\si})_i=-a^*(k_i (f_{p,\si}-f_p))-a( k_i (f_{p,\si}-f_p) ) +\big( \lan f_{p,\si}, k_i f_{p,\si}\ran  -  \lan f_{p}, k_i f_{p}\ran  \big)
+(\nabla E_{p,\si}-\nabla E_p)_i.
\eeqa
Considering that  $(\Ga_{p}^2-\Ga_{p,\si}^2)=-\Ga_p\cdot \De\Ga_{p,\si}-\De\Ga_{p,\si} \cdot \Ga_p -(\De\Ga_{p,\si})^2$,
{\crr Lemmas~\ref{dGa-lemma}  and {\ref{f-p-difference-lemma}} give}
\beqa
\|(H_{p}^{\mrm{w}})^{\ell-1} (\Ga_{p}^2-\Ga_{p,\si}^2)  \phi_{p,\si}\|_{\mcF}\leq c\si^{1/4-\de_{\la_0}}.
\eeqa
This concludes the proof of (\ref{shift-Hamiltonian-two}).  Now (\ref{shift-Hamiltonian-two-zero}) follows from Lemma~\ref{energy-bounds-w}.   \qed
\bel\label{dGa-lemma} Let $h_1, \ldots, h_{\ell}$ be real-valued measurable functions (in momentum space) which are bounded on compact sets. Then
\beqa
\|(\dGa(h_1)\ldots \dGa(h_{\ell}))\phi_{p,\si}\|_{\mcF}\leq \fr{ c_{\ell} }{\si^{\de_{\la_0}}}\big(\sup_{|k_1|\leq \ka_*} |h_1(k_1)| \ldots \sup_{|k_{\ell}|\leq \ka_*} |h_{\ell}(k_{\ell})|\big), \label{dGa-first-estimate}
\eeqa
 where $c_{\ell}$, $\de_{\la_0}$ may depend on $\ell$. Furthermore, if $f_1, \ldots, f_{{\crr \ti\ell}}\in L^2(\real^3_k)$
 are  supported in a ball of radius $\ka_*$, then we get
\beqa
\|{\crr \dGa(h_1)\ldots \dGa(h_{\ell}) a^{(*)}(f_1)\ldots    a^{(*)}(f_{\ti\ell}) }\phi_{p,\si}\|_{\mcF}\leq \fr{ c_{\ell, \ti \ell} }{\si^{\de_{\la_0}}}\big(\sup_{|k_1|\leq \ka_*} |h_1(k_1)| \ldots \sup_{|k_{\ell}|\leq \ka_*} |h(k_{\ell})|\big) \,
\big(\|f_1\|_2\ldots \|f_{\ti \ell}\|_2  \big), \label{dGa-second-estimate}
\eeqa
where  $c_{\ell,\ti\ell}$ and $\de_{\la_0}$ may depend on $\ell$, $\ti\ell$. {\crr The estimate also holds for
an arbitrary permutation of the   $(\ell+\ti\ell)$-element set of  operators on the l.h.s. of (\ref{dGa-second-estimate}). } 
\eel
\proof We consider (\ref{dGa-first-estimate}) for $\ell=1$.
Let $f^{\m}_{\mrm{w},p,\si}$ be the $\m$-photon wave functions of $\phi_{p,\si}$. Then,
by \cite[Proposition A.4]{DP19}, we have
\beqa
|f^{\m}_{\mrm{w},p,\si}(k_1,\ldots, k_{\m})|\leq \fr{1}{\sqrt{\m!}} g^{\m}_{\si}(k_1,\ldots,k_{\m}), 
\eeqa
where $g_{\si}^{{\crr n}}$ are defined as in (\ref{g-function}). Thus
\beqa
\|\dGa(h_1)\phi_{p,\si}\|^2_{\mcF}\!\!&\leq&\!\!  \sum_{\m=1}^{\infty}\fr{1}{\m!}\int d^{3{\m}}k\, (h_1(k_1)+\cdots+h_{1}(k_{\m}))^2
 |g^{\m}_{\si}(k_1,\ldots,k_{\m})|^2\non\\
\!\!&\leq&\!\!  (\sup_{|k|\leq \ka_*} |h(k)|)^2 \sum_{\m=1}^{\infty}\fr{{\m}^2}{\m!}\int d^{3\m}k\, 
 |g^{\m}_{\si}(k_1,\ldots,k_{\m})|\leq \fr{ c }{\si^{\de_{\la_0}}}(\sup_{|k|\leq \ka_*} |h(k)|)^2,
 \eeqa
where we estimated as in (\ref{N-computation}). Generalization   to arbitrary  $\ell$ is straightforward. 

As for (\ref{dGa-second-estimate}), we first commute all the operators $a^{(*)}(f_j)$ to the left and thus get
a linear combination of terms of the form
\beqa
\| a^{(*)}(f'_1) \ldots  a^{(*)}(f'_{\ti\ell})(1+N)^{-\ti\ell}(1+N)^{\ti\ell }  (\dGa(h_{i_1})\ldots   \dGa(h_{i_{\ell'}}))\phi_{p,\si}\|_{\mcF}, \label{long-expression}
\eeqa
where $f_i'(k)=h_{j_1}(k)\ldots h_{j_{l}}(k) f_i(k)$ and the functions $h_j$ included in $f_i'$ do not appear
in the product of $\dGa(h_{i_1})$ in (\ref{long-expression}). Now using the number bounds (\ref{number-bounds-eq}) on creation and annihilation operators, assumption on the supports of $f_i$ and (\ref{dGa-first-estimate}),
we obtain the claim. \qed 
\bel\label{energy-bounds-w} For any $\ell\in \nat$, the operators  $H_{\mrm{f}}^{\ell} (i+H^{\mrm{w}}_p)^{-\ell}$
are bounded.
\eel
\proof Let $\psi\in \mcF$, $\|\psi\|_{\mcF}=1$, be in the domain of $H_{\mrm{f}}^{\ell}$. Then we can write
\beqa
\|(1+H^{\mrm{w}}_p)^{-\ell} H_{\mrm{f}}^{\ell}\psi\| \!\!&\leq&\!\! \| \big( (1+H^{\mrm{w}}_p)^{-\ell}-  
(1+H^{\mrm{w}}_{p,\si})^{-\ell} \big) H_{\mrm{f}}^{\ell}\psi\|_{\mcF}\non\\
& &+ \|(1+H_{p,\si})^{-\ell}  (W(f_{p,\si})^*H_{\mrm{f}}W(f_{p,\si})    )^{\ell}\|_{\mcF}. 
\eeqa
Exploiting the concrete expression for $W(f_{p,\si})^*H_{\mrm{f}}W(f_{p,\si})$ and standard 
energy bounds for the Hamiltonian $H_{p,\si}$, i.e. the boundedness of $(1+H_{p,\si})^{-\ell}H_{\mrm{f}}^{\ell}$
(cf. \cite[Appendix D]{FGS01}), we obtain that the last term is uniformly bounded in $\si$. Now since
$\lim_{\si\to 0}H^{\mrm{w}}_{p,\si} =  H^{\mrm{w}}_{p}$  in the norm-resolvent sense, we complete the proof
by first taking $\si\to 0$ on the r.h.s. and then taking supremum over $\psi$. {\bc (The statement about the norm-resolvent
convergence is verified using the resolvent identity and explicit formulas for $H_{p,\si}^{\mrm{w}}-  H_{p}^{\mrm{w}}$,
appearing in the proof of Lemma~\ref{Hamiltonian-lemma}).} \qed

\section{Proof of Lemma~\ref{stationary-phase}}\label{A-appendix}
\setcounter{equation}{0}

Let $V:=\{\, \nabla \mcE_{p} \,|\, p\in \supp\, {\bcc g} \}$ and $V_{\de}$ be its slightly larger neighbourhood.
Since $|\nabla \mcE_p| <c_0<1$ for $p\in S$, we can ensure that $V_{\de}$ is in the interior of the  ball
of radius $c_0$ centered at zero. 
Arguing by the non-stationary phase method (cf. Theorem XI.14 from \cite{RS3} and its Corollary) 
we obtain for $x/t\notin V_{\de}$ and any $\psi$ of norm one from the dense domain in the statement of the theorem:  
\beqa
| \chi_{  \{  (x/t) \notin V_{\de} \}  }(x) \int d^3p \, e^{i(p\cdot x-\mcE_{p}t) } \lan \psi, \uu_{(t,x)}(p)\ran_{\hilb} |\leq  c(1+|x|+|t|)^{-2}   \chi_{  \{  x/t \notin V_{\de} \}  }(x)  
\sum_{|\al|\leq 2 } \sup_{p}|  \lan \psi, \pa_{p}^{\al} \uu_{(t,x)}(p)  \ran_{\hilb}|,
\eeqa
{\bcc where we write the $(t,x)$-dependence of $g$ explicitly.} 
Hence, considering that $|x|/t \geq c_0$ implies $(x/t) \notin V_{\de}$, we obtain for any $0<\eps<1$,
\beqa
& &\int_{|x|\geq c_0t}\, d^3x\,\sup_{\|\psi\|\leq 1}  \big|\lan \psi,  \int d^3p \, e^{i(p\cdot x-\mcE_{p}t) } \uu_{(t,x)}(p)\ran_{\hilb}|^2\non\\
& &\ph{44444444444444444}\leq \int_{|x|\geq c_0t} d^3x\,   c(1+|x|)^{-4+2\eps } \bigg( \sum_{|\al|\leq 2 } \sup_{p, |x'|\geq c_0t} \fr{1}{(1+|x'|+|t|)^{\eps} } \|   \pa_{p}^{\al} \uu_{(t,x')}(p) \|^2_{\hilb}\bigg)^2,
\eeqa
which gives (\ref{stationary-phase-large-vel}).

Next, by the the stationary phase method (cf. Theorem XI.15 from \cite{RS3}  and its Corollary) we have for  all $x\in \real^3$
\beqa
\big| \int d^3p \, e^{i(p\cdot x-\mcE_{p}t) }\lan \psi, \uu_{(t,x)}(p)\ran_{\hilb} \big|\leq ct^{-3/2}   \sum_{|\al|\leq 2 }  \sup_{p} |\lan\psi, \pa_{p}^{\al} \uu_{(t,x)}(p)\ran_{\hilb}|,
\label{stationary-phase-estimate}
\eeqa 
where we used that the Hessian matrix of $p\mapsto \mcE_p$  
 is invertible in $S$ (cf. Lemma~\ref{energy-lemma}). Thus we get
\beqa
\int_{|x|\leq c_0t}\, d^3x\,\sup_{\|\psi\|_{\mcF}\leq 1}  \big|\lan \psi,   \int d^3p \, e^{i(p\cdot x-\mcE_{p}t) } \uu_{(t,x)}(p)\ran_{\hilb}|^2
\leq  \int_{|x|\leq c_0t}d^3x\,  c(t^{-3} ) \bigg(\sum_{|\al|\leq 2 }   \sup_{p, |x'|\leq c_0t}
\|\pa_{p}^{\al} \uu_{(t,x')}(p) \|_{\hilb} \bigg)^2,
\eeqa 
which gives (\ref{stationary-phase-small-vel}). 

{\bc To prove that the constants $c$ are uniform in $|\la|\in (0,\la_0]$,  we use the last statement 
in  \cite[Theorem XI.15]{RS3} and argue as in the corollary of this theorem. 
This argument requires, that  $\la\mapsto \{E_p^{(\la)}\}_{p\in S}$ is continuous in the topology of $C^{\ell}(S)$ (cf. \cite[p.37]{RS3}). This follows from Lemma~\ref{energy-lemma}}. \qed 

\section{Properties of functions $f_p, {\rcc f_{p,\si}}$}
\setcounter{equation}{0}

{\crr We recall from (\ref{Bogolubov}), (\ref{Bogolubov-cut}) the definitions:
\beqa
f_p(k):=\la\fr{\chi_{\ka}(k)}{\sqrt{2|k|}}\fr{1}{|k|(1-e_k\cdot \nabla E_{p} )}, \quad
f_{p,\si}(k):=\la\fr{\chi_{[\si,\ka)}(k)}{\sqrt{2|k|}}
\fr{1}{|k|(1-e_k\cdot \nabla E_{p,\si} )}. \label{f-def}
\eeqa
We start with the following preparatory lemma. 
}
\bel\label{f-p-lemma} For $n=1,2,\ldots$ there holds the bound
\beqa
 \int   \fr{\chi_{\ka}(k)^2 }{ |k|^{3} }  |e^{-i|k|  t+ik\cdot x }-1|^n d^3k\leq c_n  (1+\log(1+|t|+|x|)).
\eeqa
\eel
\proof We estimate
\begin{align}
 \int   \fr{\chi_{\ka}(k)^2 }{ |k|^{3} }  |e^{-i|k|  t+ik\cdot x }-1|^n d^3k
&\leq  \sum_{\eps=\pm} \int_{ \eps(t -e_k\cdot x)\geq 0  } d\Om(e_k) \int_0^{\ka} \fr{d|k|}{|k|} \,  |e^{-i\eps  |k| \eps (t -e_k\cdot x ) }-1|^n\non\\
&\leq  \sum_{\eps=\pm} \int_{ \eps(t -e_k\cdot x)\geq 0  } d\Om(e_k) \int_0^{\ka  (1+ |t|+|x| ) } \fr{d|k|}{|k|} \,  |e^{-i\eps  |k|  }-1|^n\non\\
&\leq  c  \int_0^{\ka } \fr{d|k|}{|k|} \,  |e^{-i  |k|  }-1|^n+ c {\bcc 2^{n}}\int_{\ka}^{\ka  (1+ |t|+|x| ) } \fr{d|k|}{|k|}\non\\
&\leq c_n(1+\log(1+|t|+|x|)). \label{integral-lemma}
\end{align}
This completes the proof. \qed
\bel\label{theta-Im-lemma} There holds the following bound for $|\al|=0,1,2$
\beqa
& &|\pa_p^{\al}\thet(p,x,t)|\leq  c|\la|^2(1+\log(1+|t|+|x|)). \label{f_p-zero-estimate}
\eeqa
\eel
\proof We have
\begin{align}
\pa_{p_i}f_{p}(k)&=\la \fr{ \chi_{\ka}(k)}{ \sqrt{2}  |k|^{3/2} }  \fr{   1    }{(1-e_k\cdot \nabla E_{p} )^2}   \pa_{p_i}(e_k\cdot \nabla E_{p}),
\label{f-comp-one}\\
\pa_{p_j}\pa_{p_i}f_{p}(k)&=\la \fr{ \chi_{\ka}(k)}{ \sqrt{2}  |k|^{3/2} } \bigg\{ 2\fr{   1    }{(1-e_k\cdot \nabla E_{p} )^3} \pa_{p_j}(e_k\cdot \nabla E_{p}) 
 \pa_{p_i}(e_k\cdot \nabla E_{p}) \non\\ & \ph{4444444444444444444}+  \fr{   1    }{(1-e_k\cdot \nabla E_{p} )^2}   \pa_{p_j}\pa_{p_i}(e_k\cdot \nabla E_{p}) \bigg\}. \label{f-comp-two}
\end{align}
Thus by Lemma~\ref{energy-lemma}, we have
\beqa
|f_{p}(k)|, \,  |\pa_{p_i}f_{p}(k)|, \,   {\bcc | \pa_{p_j}\pa_{p_i}f_p(k)|} \leq  c\la \fr{ \chi_{\ka}(k)}{ \sqrt{2}  |k|^{3/2} }, 
 \label{f-p-estimates}
\eeqa
Now we write
\begin{align}
\pa_p^{\al}\thet(p,x,t)&= \int d^3k\, \pa_p^{\al}(f_p(k)^2)  \sin(|k|t- k\cdot x ).
\end{align}
Clearly, estimates~(\ref{f-p-estimates}) and Lemma~\ref{f-p-lemma} give (\ref{f_p-zero-estimate}). 
(Here we made use of $|\sin\, y|=|\mrm{Im}\, e^{i y}|=|\mrm{Im} (e^{iy}-1)|\leq |e^{i y}-1|$). \qed
\bel\label{F-lemma} Let $m(t,x):= (e^{-i|k|  t+ik\cdot x }-1)$. Then, for $|\al|\leq 2, |\be|\leq 2$
\beqa
 {\crr |} \lan  \pa_{p}^{\al} f_{p}m(t,x), \pa_{p}^{\be}f_{p}m(t,x)  \ran {\crr |}  \leq   c|\la|^2 (1+\log(1+|t|+|x|)).   
\eeqa 
\eel
\proof  Follows immediately from (\ref{f-p-estimates}) and Lemma~\ref{f-p-lemma}. Indeed, we have
\beqa
\int d^3k\, {\crr |}  (\pa_{p}^{\al} f_p)(k)   (\pa_{p}^{\be} f_p)(k)  {\crr |}\,    | e^{-i|k|  t+ik\cdot x }-1|^2
\leq  c|\la|^2 \int d^3k\,  \fr{ \chi_{\ka}(k)^2   }{ 2  |k|^{3} }     | e^{-i|k|  t+ik\cdot x }-1|^2\leq c|\la|^2  (1+\log(1+|t|+|x|)),\,\,
\eeqa
which concludes the proof. \qed
{\mg 
\bel\label{new-lemma} The following bounds hold 
\beqa
{\crr |} \pa_p^{\al} f_{p,\si}(k) {\crr |} \leq   c \la \fr{ \chi_{[\si, \ka)}(k)}{ \sqrt{2}  |k|^{3/2} } \ \ \mathrm{for} \ \  |\al|=0,1 \ \ \mathrm{and} \ \  {\crr |}\pa_p^{\al} f_{p,\si}(k) {\crr |} \leq  \fr{c}{\si^{\de_{\la_0}}}\la \fr{ \chi_{[\si, \ka)}(k)}{ \sqrt{2}  |k|^{3/2} } \ \ \mathrm{for} \ \  |\al|=2.
\eeqa
\eel
\proof The estimates follow from definition~(\ref{Bogolubov-cut}) via computations analogous to (\ref{f-comp-one})--(\ref{f-comp-two}).
For the relevant estimates on derivatives of $S\ni p\mapsto E_{p,\si}$ up to the third order, see \cite[Theorem 2.1]{DP12}.  \qed
}
{\crr
\bel \label{f-p-difference-lemma} The following bounds hold
\beqa
\|k_i(f_{p,\si}-f_p)\|_2\leq c\si^{1/4}, \quad |\lan f_{p,\si}, k_i f_{p,\si}\ran - \lan f_{p}, k_i f_{p}\ran| \leq c\si^{1/4}.
\label{two-f-bounds}
\eeqa
\eel
\proof Definitions (\ref{f-def}) give
\beqa
f_{p}(k)-f_{p,\si}(k)=\la\fr{\chi_{[0,\si]}(k) }{\sqrt{2}|k|^{3/2}} \fr{1}{(1-e_k\cdot \nabla E_p)}
+\la \fr{\chi_{[\si,\ka]}(k)  }{\sqrt{2} |k|^{3/2}} \bigg( \fr{1}{(1-e_k\cdot \nabla E_p)}-  \fr{1}{(1-e_k\cdot \nabla E_{p,\si})}\bigg),
\eeqa
where $\chi_{[0,\si]}$ is the characteristic function of a ball of radius $\si$ centered at zero. Now considering that 
$|\nabla E_{p}-\nabla E_{p,\si}|\leq c\si^{1/4}$ (cf. (\ref{standard-bounds}) above), we can write for any $\be>0$ 
\beqa
\| \,|k|^{\be}(f_{p,\si}-f_p)\|_2\leq c(\si^{\be}+\si^{1/4}). \label{beta-estimate}
\eeqa
Setting $\be=1$ we obtain the first bound in (\ref{two-f-bounds}). As for the second estimate, 
we note that
\beqa
|\lan f_{p,\si}, k_i f_{p,\si}\ran - \lan f_{p}, k_i f_{p}\ran| \!\!&\leq&\!\! \int d^3k\,|k| \bigg| (f_{p,\si}(k)-  f_{p}(k))(f_{p,\si}(k)+ f_{p}(k))
 \bigg|\non\\
\!\!&\leq&\!\! \|\,|k|^{1/2} (f_{p,\si}-  f_{p})\|_2 \big( \|\, |k|^{1/2} f_{p,\si} \|_2+ \|\,|k|^{1/2} f_{p} \|_2\big). 
\eeqa
Applying (\ref{beta-estimate}) with $\be=1/2$ and considering that $\|\, |k|^{1/2} f_{p,\si} \|_2, \|\,|k|^{1/2} f_{p} \|_2\leq c$
we conclude the proof. \qed 
}



\end{document}